\documentclass[prb,groupedaddress,superscriptaddress,reprint,twocolumn,notitlepage]{revtex4}

\usepackage{amsfonts}
\usepackage{amsmath}
\usepackage{amssymb}
\usepackage{graphicx}
\usepackage{hyperref}
\usepackage{color}
\usepackage{mathrsfs}
\usepackage{bbm}
\usepackage{subfigure}
\usepackage{palatino}

\newcommand{\ignore}[1]{}
\newcommand{\nobibentry}[1]{{\let\nocite\ignore\bibentry{#1}}}

\newcommand{\ket}[1]{\left\vert#1\right\rangle}
\newcommand{\bra}[1]{\left\langle#1\right\vert}

\newcommand{\eket}[2]{\left\vert #1_{\text{#2}}\right\rangle}
\newcommand{\ebra}[2]{\left\langle #1_{\text{#2}}\right\vert}
\newcommand{\meket}[2]{\left\vert #1_{#2}\right\rangle}
\newcommand{\mebra}[2]{\left\langle #1_{#2}\right\vert}
\newcommand{\eeket}[4]{\left\vert #1_{\text{#3}} #2_{\text{#4}}\right\rangle}
\newcommand{\eebra}[4]{\left\langle #1_{\text{#3}} #2_{\text{#4}} \right\vert}
\newcommand{\eeeket}[6]{\left\vert #1_{\text{#4}} #2_{\text{#5}} #3_{\text{#6}} \right\rangle}
\newcommand{\eeebra}[6]{\left\langle #1_{\text{#4}} #2_{\text{#5}} #3_{\text{#6}} \right\vert}

\DeclareMathAlphabet{\mathpzc}{OT1}{pzc}{m}{it}

\begin{document}

\title{Quantum-enhanced absorption refrigerators}

\author{Luis A. Correa}
\affiliation{IUdEA Instituto Universitario de Estudios Avanzados, Universidad de La Laguna, La Laguna 38203, Spain}
\affiliation{Dpto. F\'{\i}sica Fundamental, Experimental, Electr\'{o}nica y
Sistemas, Universidad de La Laguna, La Laguna 38203, Spain}
\affiliation{School of Mathematical Sciences, The University of Nottingham, University Park, Nottingham NG7 2RD, UK}

\author{Jos\'{e} P. Palao}
\affiliation{IUdEA Instituto Universitario de Estudios Avanzados, Universidad de La Laguna, La Laguna 38203, Spain}
\affiliation{Departamento de F\'{i}sica Fundamental II, Universidad de La Laguna, La Laguna 38204, Spain}

\author{Daniel Alonso}
\affiliation{IUdEA Instituto Universitario de Estudios Avanzados, Universidad de La Laguna, La Laguna 38203, Spain}
\affiliation{Dpto. F\'{\i}sica Fundamental, Experimental, Electr\'{o}nica y
Sistemas, Universidad de La Laguna, La Laguna 38203, Spain}

\author{Gerardo Adesso}
\affiliation{School of Mathematical Sciences, The University of Nottingham, University Park, Nottingham NG7 2RD, UK}

\date{\today}

\maketitle

{\bf Thermodynamics is a branch of science blessed by an unparalleled combination of generality of scope and formal simplicity. Based on few natural assumptions together with the four laws, it sets the boundaries between possible and impossible in macroscopic aggregates of matter. This triggered groundbreaking achievements in physics, chemistry and engineering over the last two centuries. Close analogues of those fundamental laws are now being established at the level of individual quantum systems, thus placing limits on the operation of quantum-mechanical devices. Here we study quantum absorption refrigerators, which are driven by heat rather than external work. We establish thermodynamic performance bounds for these machines and investigate their quantum origin. We also show how those bounds may be pushed beyond what is classically achievable, by suitably tailoring the environmental fluctuations via quantum reservoir engineering techniques. Such superefficient quantum-enhanced cooling realises a promising step towards the technological exploitation of autonomous quantum refrigerators.}
\newline

An absorption or heat-driven quantum refrigerator is a system capable of establishing a net steady-state transport of energy from a cold bath (c) to a hot bath (h), assisted only by the residual heat coming from an additional work reservoir (w) \cite{PhysRevE.64.056130,PhysRevLett.105.130401,PhysRevLett.108.070604}. In this picture, the cold bath would play the role of the macroscopic or mesoscopic object to be cooled. In addition to their potential technological applications, these autonomous quantum-thermal devices are also appealing from the fundamental perspective, as they are naturally well suited for the study of thermodynamics at the level of individual open quantum systems \cite{geva1996quantum,PhysRevE.64.056130,PhysRevE.85.051117,PhysRevE.85.061126}.

In spite of the increasing interest that quantum absorption cooling has attracted over the last few years \cite{1751-8121-44-49-492002,PhysRevE.85.051117,PhysRevE.87.042131,0295-5075_97_4_40003,PhysRevLett.110.256801,1305.6009v1,PhysRevLett.108.120602}, the field is far from new. A heat-driven quantum fridge is just one specific configuration of the more general quantum heat pump, that can function either as a heater, a chiller or even an engine. The use of three-level solid-state \textit{masers} as physical support for heat pumps was already discussed in the late 1950s \cite{PhysRevLett.2.262,PhysRev.156.343}, when spin refrigeration was also experimentally demonstrated \cite{5123190}. The consistent quantum-thermodynamic description of these elementary three-level prototypes was object of further study \cite{geva1996quantum,PhysRevE.64.056130} and, just recently, alternative finite-dimensional quantum systems realising autonomous heat pumps have been put forward in the literature \cite{PhysRevLett.105.130401,PhysRevLett.108.070604}.

The different designs of quantum heat pumps share limitations that can be understood from the assumptions on their interactions with the environments. Under the familiar conditions usually met in the quantum-optical regime, the dissipative processes may be assumed purely Markovian \cite{lindblad1976generators,gorini1976completely,spohn1978entropy}, which severely restricts the performance of any heat-driven device, and confers a distinctive spectral structure to the environmental fluctuations \cite{breuer2002theory}. In particular, once their steady state builds up, quantum heat pumps are governed by formal analogues of the laws of thermodynamics and, as a consequence, their absolute efficiency ideally saturates to the corresponding Carnot limits $\varepsilon_C$, albeit at vanishing `cooling power' \cite{PhysRev.156.343}, i.e. in the reversible limit, the exchange of any finite amount of energy with the heat baths is performed in infinite time.

For practical purposes, however, one needs to operate at nonvanishing power. In this case, the relevant issue to assess the functionality of these devices demands the optimisation of more practical figures of merit such as the efficiency at maximum cooling power $\varepsilon_*$. The natural question arises whether $\varepsilon_*$ can approach $\varepsilon_C$ arbitrarily closely even at finite cooling power, or if, on the contrary, it is upper bounded by some fundamental limit. The efficiency at maximum `mechanical' power is extensively used to benchmark the operation of heat engines and a lot of effort has been devoted to establish a universal upper bound therefor \cite{curzon1975efficiency,PhysRevLett.95.190602,PhysRevLett.105.150603}. Unfortunately, the general arguments used for engines do not provide simple bounds when the cycle is reversed into a refrigerator, and consequently, a different approach is needed to arrive to model-independent performance bounds. Here, we rigorously prove that the `smallest' quantum absorption refrigerators, supported on ideal three-level masers, are limited in their efficiency at maximum power by a fraction of $\varepsilon_C$, only related to the spectral properties of the environmental fluctuations at low frequencies. We show that this general performance bound applies as well to `larger' and non-ideal designs \cite{PhysRevLett.105.130401,PhysRevLett.108.070604,PhysRevE.87.042131}, as it is independent of the details of the working material of the refrigerator.

Achieving a good understanding of the quantum-mechanical origin of the limitations of heat pumps can also provide key clues about how to surmount them. We show indeed that, by feeding an absorption fridge with engineered thermal resources, one can push its performance bounds considerably further, allowing for classically impossible superefficient quantum cooling. Namely, at given fixed environmental temperatures, the addition of squeezing to the work bath leads to efficiencies above $\varepsilon_C$ and, most interestingly, to a systematic enhancement of the output harnessed power. This is achieved strictly within the framework of quantum thermodynamics and thus, in no violation of its laws \cite{e15062100,PhysRevE.85.061126}.

\section*{Results}

\begin{figure*}[t]
	\subfigure{\label{Fig1a}
	\includegraphics[width=0.32\textwidth]{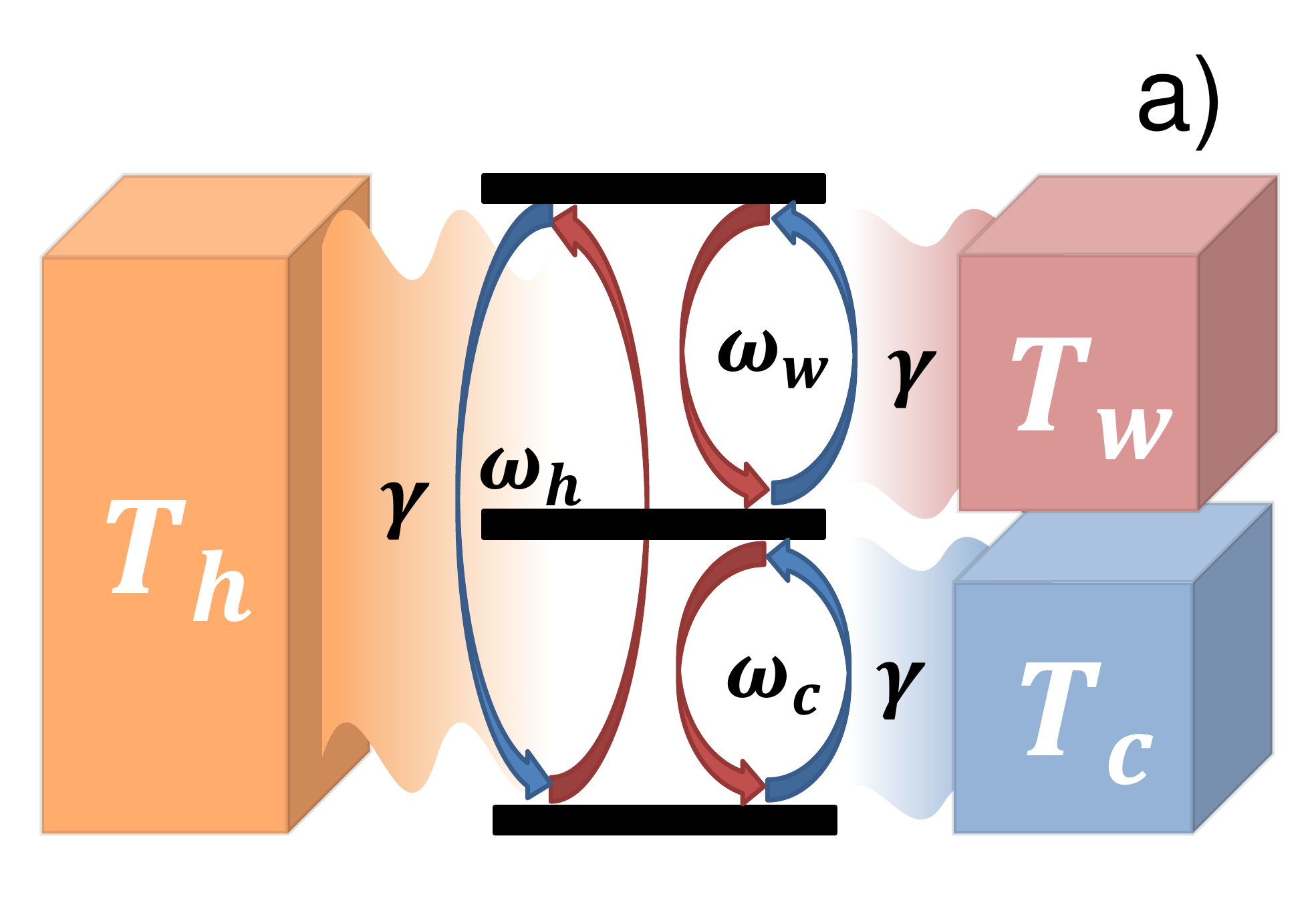}}
	\subfigure{\label{Fig1b}
	\includegraphics[width=0.32\textwidth]{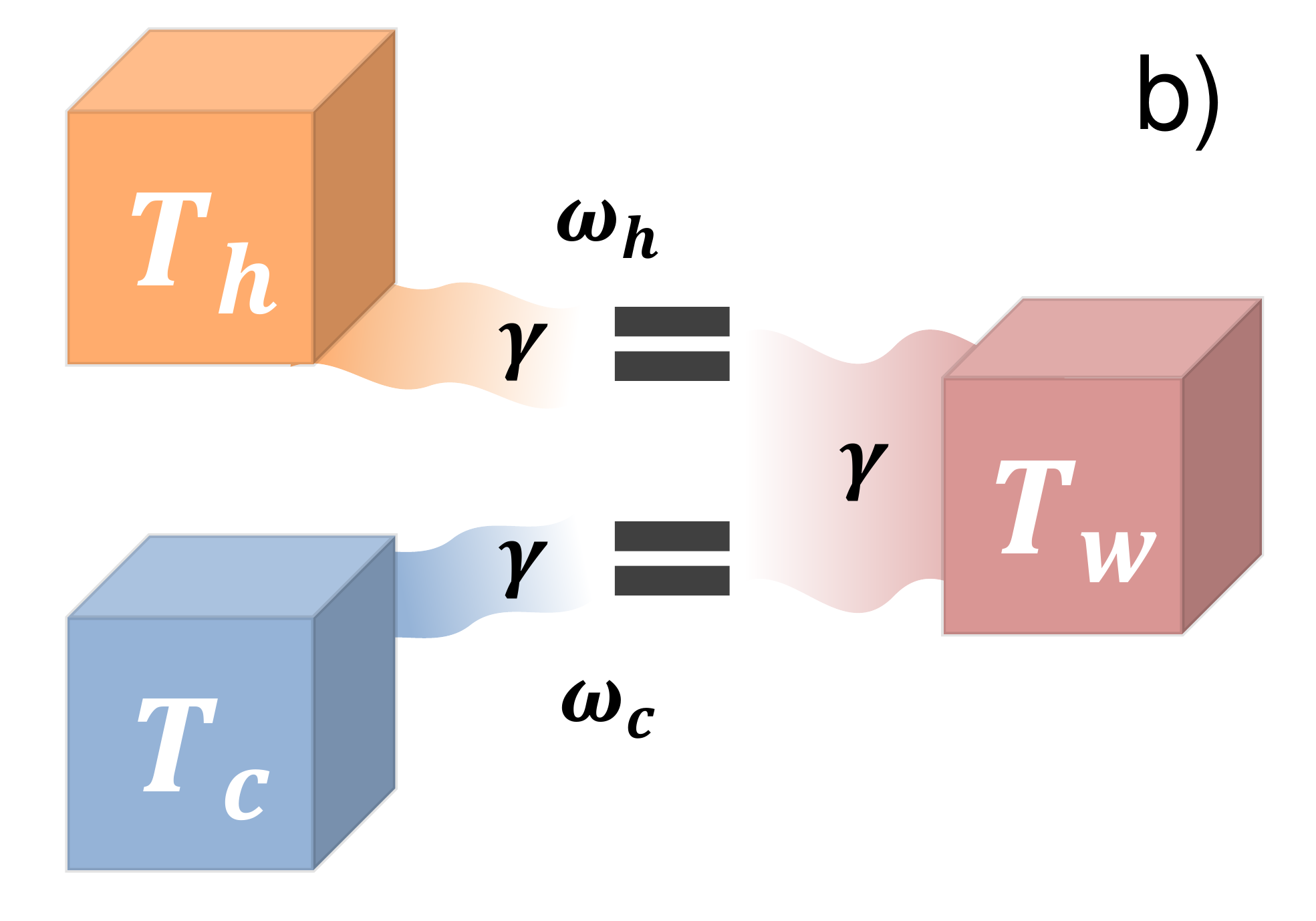}}
	\subfigure{\label{Fig1c}
	\includegraphics[width=0.32\textwidth]{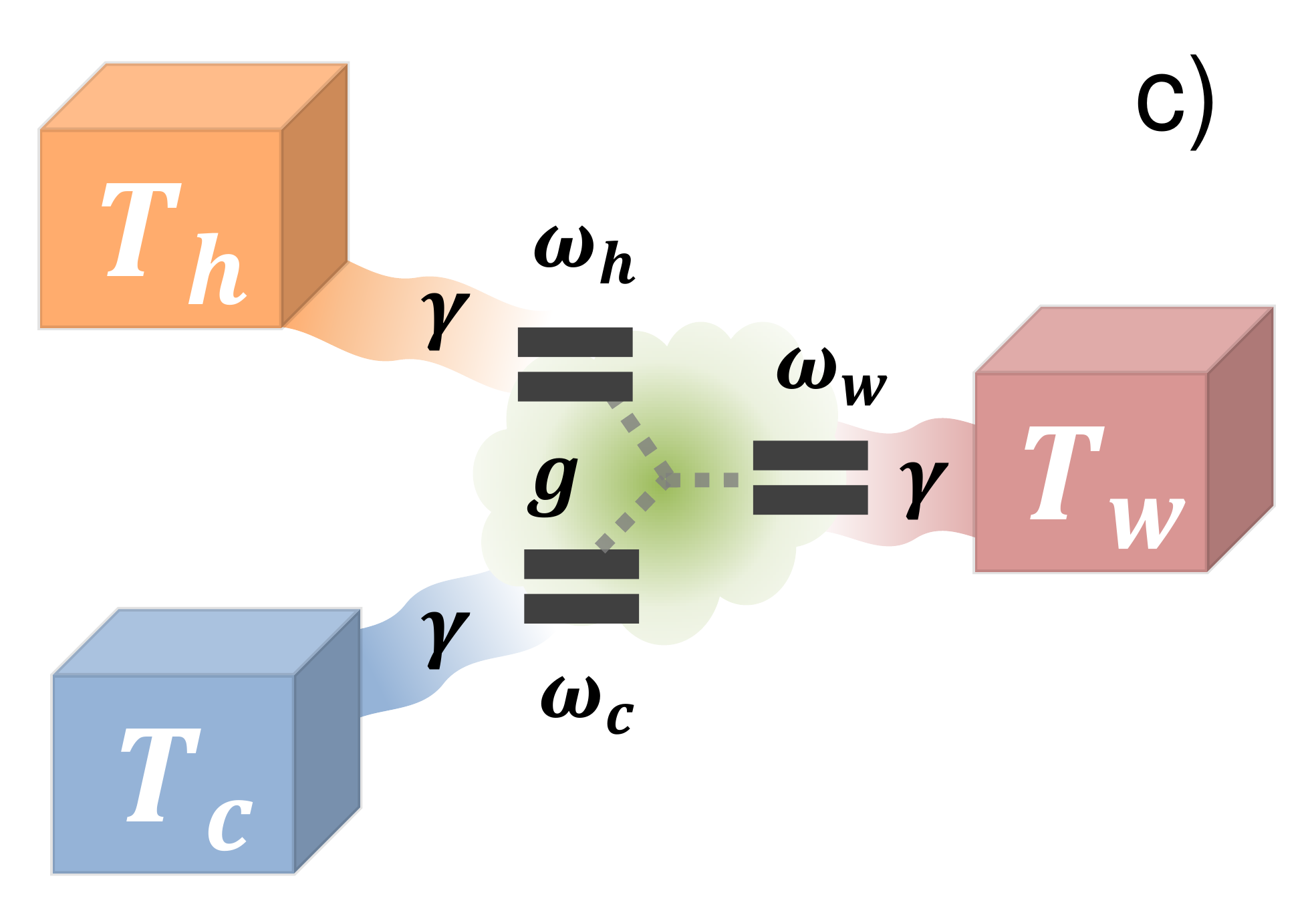}}
\caption{(a) Schematic representation of the three-level heat pump. The three independent baths are taken as infinite collections of uncoupled modes with Hamiltonian $\hat H_{B_\alpha}=\sum_\mu \omega_\mu \hat b_{\alpha,\mu}^\dagger \hat b_{\alpha,\mu}$. The dissipative interactions in each bath-transition pair are modeled by terms like e.g. $\hat H^\text{int}_c=\sqrt{\gamma}(\ket{0}\bra{1}+\ket{1}\bra{0})\otimes\hat{\mathcal{B}}_c$, with $\hat{\mathcal{B}}_\alpha=\sum_\mu g_{\alpha,\mu} (\hat b_{\alpha,\mu}+ \hat b_{\alpha,\mu}^\dagger)$. Here, the coupling constants $g_{\alpha,\mu}$ are proportional to $\sqrt{\omega_\mu}$, which results in flat spectral densities, and the parameter $\gamma$ controls the overall strength of the interaction \cite{breuer2002theory}. The three-stroke cooling and heating processes are depicted with blue and red arrows respectively. (b) Diagram of a two-qubit heat pump, where we shall always assume $\omega_h>\omega_c$. Energy transport between the two contacts is mediated by dissipation into the work bath thanks to a coupling term of the form $\hat H^\text{int}_w=(\eket{1}{c}\ebra{0}{h}+\eket{0}{c}\ebra{1}{h})\otimes\hat{\mathcal{B}}_w$ \cite{PhysRevLett.108.070604}. (c) A design of heat pump with a contact qubit for each bath. In this case, the working material is provided with an explicit three-body interaction term to allow for energy exchanges $\hat{H}_\text{wm}=\hat{H}_0+\hat{H}_I$, with $\hat{H}_I=g(\eeeket{1}{0}{1}{w}{h}{c}\eeebra{0}{1}{0}{w}{h}{c}+\text{h.c.})$. The frequencies are constrained by $\omega_h=\omega_c+\omega_w$. \cite{PhysRevLett.105.130401}}
\label{Fig1}
\end{figure*}

\noindent\textbf{Models of absorption refrigerator.}
As already advanced, a minimal model of autonomous heat pump \cite{PhysRevLett.2.262} consists of a three-level system with each of its transitions weakly coupled to one of the three independent heat baths [see Fig.~\ref{Fig1}(a)]. Essentially, as the steady state builds up, the `heat' collected from the cold bath is dumped into the hot bath with the assistance of the extra energy provided by the work bath, which closes the cooling cycle. Of course, the opposite heating cycle also takes place in the steady-state, and it is the imbalance between these two stationary processes which renders the device either a refrigerator or a heater. As we shall see below, refrigeration occurs as long as the frequency of the transition coupled to the cold bath remains below a certain threshold $\omega_c\leq\omega_{c,\max}$ \cite{PhysRevE.64.056130}.

We shall also consider the two-qubit design \cite{PhysRevLett.108.070604,PhysRevE.85.061126} of Fig.~\ref{Fig1}(b) in which the cold and the hot bath are each addressed through a two-level contact, while the work reservoir introduces dissipation in the subspace $\{\left\vert 0_\text{c} \right\rangle\otimes\left\vert 1_\text{h} \right\rangle,\left\vert 1_\text{c} \right\rangle\otimes\left\vert 0_\text{h} \right\rangle\}$, where $\left\vert 0_{\alpha} \right\rangle$ and $\left\vert 1_{\alpha} \right\rangle$ stand for the ground and excited states of contact $\alpha\in\{\text{h},\text{c}\}$. This allows for energy flows between the hot and cold ends, and eventually results in net refrigeration. Remarkably, this fridge works within the same cooling window ($\omega_c\leq\omega_{c,\max}$) as the three-level prototype.

Finally, we shall comment on another model \cite{PhysRevLett.105.130401,1751-8121-44-49-492002,PhysRevE.85.051117} featuring three contact qubits connected via a three-body interaction [see Fig.~\ref{Fig1}(c)]. Practically-oriented issues have been recently studied in connection with this design, including its potential experimental realisations \cite{0295-5075_97_4_40003,PhysRevLett.110.256801} and the investigation of its efficiency at maximum power \cite{PhysRevE.87.042131}.
\newline

\noindent\textbf{The first and second laws.}
When the interaction of the working material with the environments is sufficiently weak, one can tackle its effective dynamics via a quantum master equation like
\begin{equation}
\frac{d}{dt}~\hat\varrho(t)= \left(\mathcal{D}_w + \mathcal{D}_h + \mathcal{D}_c\right) ~\hat\varrho(t).
\label{lindblad}
\end{equation}
It is an equation of motion for the reduced state of the heat pump $\hat \varrho(t)$, where $\mathcal{D}_\alpha$ are the dissipation super-operators associated with each bath. An initial preparation $\hat\rho(0)=\hat\varrho(0)\bigotimes_\alpha\hat\chi^{T}_\alpha$, factorised between system and baths degrees of freedom is assumed. The intrinsic dynamics has been eliminated by taking the interaction picture with respect to the free Hamiltonian of the working material $\hat H_\text{wm}$.

If the dissipation is much slower than both the environmental fluctuations and the coherent evolution of the heat pump, the Born, Markov and rotating-wave approximations may be safely applied \cite{breuer2002theory}. This leads to dissipation super-operators of the well known Lindblad-Gorini-Kossakovski-Sudarshan (LGKS) type \cite{lindblad1976generators,gorini1976completely}, which are the workhorse of quantum thermodynamics \cite{e15062100}. We shall specifically denote them by $\mathcal{L}_\alpha$ in what follows:
\begin{equation}
\mathcal{L}_\alpha\hat\varrho = \sum_{\omega} \Gamma_{\omega,\alpha} (\hat A_\alpha(\omega) \hat \varrho \hat A_\alpha^\dagger(\omega) - \frac{1}{2} \{\hat A^\dagger_\alpha(\omega)\hat A_\alpha(\omega),\hat \varrho\}_+ ).
\label{Lindblad_dissipator}\end{equation}
Here, $\{\cdot,\cdot\}_+$ stands for an anticommutator and $\hat A_{\alpha}(\omega)$, for the \textit{jump} operator associated with the decay process into channel $\omega$, which occurs at a rate given by the corresponding element of the spectral correlation tensor $\Gamma_{\omega,\alpha}$. For equilibrium reservoirs, these latter relate via the detailed balance condition \cite{breuer2002theory}
\begin{equation}
\Gamma_{-\omega,\alpha}=e^{-\hbar\omega/k_B T_\alpha}\Gamma_{\omega,\alpha}.
\label{detailed_balance}\end{equation}
Individually, each LGKS dissipator $\mathcal{L}_\alpha$ generates a completely positive and trace preserving contractive dynamics of the working material, converging towards its local stationary thermal state $\hat\tau_\alpha \propto \exp\{-\hat H_\text{wm}/k_B T_\alpha \}$ \cite{spohn1978entropy}. As time goes to infinity, this contractivity translates into (see Methods for details)
\begin{equation}
\frac{\dot{\mathcal{Q}}_w}{T_w}+\frac{\dot{\mathcal{Q}}_h}{T_h}+\frac{\dot{\mathcal{Q}}_c}{T_c} \leq 0,
\label{second_law}\end{equation}
where $T_\alpha$ are the equilibrium temperatures of the baths and $\dot{\mathcal{Q}}_\alpha=\text{tr}[\hat H_\text{wm}\mathcal{L}_\alpha\hat\varrho(\infty) ]$ represents the energy per unit time flowing from bath $\alpha$ into the pump once in the steady state. In particular, we will refer to $\dot{\mathcal{Q}}_c$ as the `cooling power' or just power.

Since $\hat{H}_\text{wm}$ is time-independent, the stationarity of the average energy at $\hat\varrho(\infty)$ implies
\begin{equation}
\dot{\mathcal{Q}}_w+\dot{\mathcal{Q}}_h+\dot{\mathcal{Q}}_c=0.
\label{first_law}\end{equation}

This balance equation plays the role of a \emph{first law} in quantum thermodynamics as soon as one identifies the steady-state currents $\dot{\mathcal{Q}}_\alpha$ with heat flows. Similarly, Eq.~\eqref{second_law} can be regarded as a quantum-thermodynamic statement of Clausius theorem, i.e. the \emph{second law}.

The combination of Eqs.~\eqref{second_law} and \eqref{first_law} places the ultimate thermodynamic bounds on the efficiency of a heat pump in its various modes of operation \cite{PhysRev.156.343}. For instance, in the chiller configuration (i.e. $\dot{\mathcal{Q}}_c>0$, $\dot{\mathcal{Q}}_w>0$ and $\dot{\mathcal{Q}}_h<0$), the efficiency is defined as the ratio of the cooling power to the input heat (per unit time) provided by the work bath, i.e. $\varepsilon\equiv\dot{\mathcal{Q}}_c/\dot{\mathcal{Q}}_w$. Using Eq.~\eqref{first_law} to eliminate $\dot{\mathcal{Q}}_h$ from Eq.~\eqref{second_law}, yields
\begin{equation}
\varepsilon=\frac{\dot{\mathcal{Q}}_c}{\dot{\mathcal{Q}}_w}\leq\frac{(T_w-T_h)T_c}{(T_h-T_c)T_w}\equiv\varepsilon_C,
\label{efficiency}\end{equation}
where $\varepsilon_C$ is nothing but the Carnot efficiency of a (macroscopic) heat driven quantum absorption fridge operating between baths at temperatures $\{T_w,T_h,T_c\}$ \cite{gordon2000cool}.
\newline

\noindent\textbf{The cooling window.}
We now want to delimit the region within the space of parameters of a quantum heat pump where cooling is permitted by the second law. This is of course highly model-dependent but one can always stick to $n$-dimensional working materials with the basic three-stroke cooling mechanism of the three-level maser built in \cite{PhysRevLett.2.262}. If the bath $\alpha$ couples to the heat pump only allowing for transitions with a gap of $\hbar \omega_\alpha$ among the eigenstates of $\hat H_\text{wm}$, and if the resonance condition $\omega_h=\omega_c+\omega_w$ holds, one should have
\begin{equation}
\left\vert\dot{\mathcal{Q}}_\alpha/\dot{\mathcal{Q}}_{\beta}\right\vert=\omega_\alpha/\omega_{\beta}.
\label{ideal}\end{equation}
This was already acknowledged as a distinctive feature of ideal three-level heat pumps in the seminal paper by Geusic et al. \cite{PhysRev.156.343}, and it is easy to see that it remains true for the two-qubit design of Fig.~\ref{Fig1}(b). Eq.~\eqref{ideal} essentially says that in a cooling cycle, every single cold excitation is traded for one hot excitation at the expense of consuming a single work excitation \cite{1751-8121-44-49-492002}.

Hence, combining Eq.~\eqref{second_law} with \eqref{ideal} yields
\begin{eqnarray}
&&\frac{\omega_w}{T_w}-\frac{\omega_h}{T_h}+\frac{\omega_c}{T_c} \leq 0 \nonumber \\
&\Longrightarrow&\omega_c\leq\omega_{c,\max}\equiv\frac{(T_w-T_h)T_c}{(T_w-T_c)T_h}\omega_h.
\label{cooling_window}\end{eqnarray}
This inequality defines the `cooling window'. Note that the work temperature $T_w$ must be larger than $T_h$ (and $T_c$) in order to have a positive $\omega_{c,\max}$.

In principle, the efficiency of any ideal heat pump satisfying Eq.~\eqref{ideal} saturates to the Carnot bound in the \emph{reversible} limit of $\omega_c\rightarrow\omega_{c,\max}$, i.e. when the contact transitions locally equilibrate with their corresponding baths so that the equality in Eq.~\eqref{second_law} holds.

On the contrary, the operation of a heat pump might become intrinsically irreversible if additional mechanisms of energy exchange were present. For instance, the three-qubit device of Fig.~\ref{Fig1}(c) only behaves as an ideal heat pump when the effects of the global interaction term $\hat H_{I}$ on the dissipative dynamics are entirely neglected [see caption of Fig.~\ref{Fig1}(a)]. Indeed, explicit analytical formulas for its steady state consistent with Eq.~\eqref{ideal} may be written down in that limit \cite{1751-8121-44-49-492002}. However, since the contact transitions are chosen among the eigenstates of $\hat H_0$ rather than those of $\hat H_\text{wm}=\hat H_0 + \hat H_I$, dissipation is always strictly `delocalised' regardless of the interaction strength, and neither Eq.~\eqref{ideal}, nor the equality in Eq.~\eqref{second_law} can be satisfied. Consequently, this specific design is an example of non-ideal heat pump, as it is unable to reach the Carnot efficiency \cite{PhysRevE.87.042131}.
\newline

\noindent\textbf{Efficiency at maximum cooling power: A model-independent bound.} In spite of its fundamental importance, the attainability of the Carnot efficiency in microscopic quantum heat pumps \cite{1751-8121-44-49-492002,1305.6009v1} is not the central issue for practical applications. Indeed, when operating at the reversible limit the power exactly vanishes.

One would like instead to run a refrigeration cycle, carefully tuning the design parameters so that efficiency and power are maximised \emph{jointly}, pretty much in the spirit of finite-time thermodynamics. The practically relevant questions would then be whether there exists a tight upper bound for the efficiency at maximum power $\varepsilon_*$ other than $\varepsilon_C$, and whether such bound is model-independent. Long-standing problems of this sort have been intensively studied in classical macroscopic heat engines and refrigerators \cite{curzon1975efficiency,PhysRevLett.78.3241,PhysRevLett.95.190602,PhysRevLett.105.150603,PhysRevE.86.011127}, as well as in their quantum-mechanical counterparts \cite{geva1991spin,PhysRevE.81.051129,PhysRevE.82.011120,abe2011power,wang2012finite,PhysRevLett.109.203006,PhysRevE.87.012140}.
In particular, a specific bound for $\varepsilon_*$ has been recently established for the model of Fig.~\ref{Fig1}(c) based on a numerical analysis, together with design prescriptions for its saturation \cite{PhysRevE.87.042131}.

In this paper, under natural assumptions on the environmental fluctuations, we prove analytically that the efficiency at maximum cooling power of any ideal fridge made up of elementary three-level heat pumps is tightly upper bounded by
\begin{equation}
\varepsilon_* \leq \frac{d_c}{d_c+1}\varepsilon_C,
\label{performance}\end{equation}
where $d_c$ stands for the spatial dimensionality of the cold bath. This applies directly to the ideal absorption refrigerators of Figs.~\ref{Fig1}(a) and \ref{Fig1}(b), but we verify that Eq.~(\ref{performance}) holds as well for the non-ideal refrigerator of Fig.~\ref{Fig1}(c), thus validating and generalising the bound obtained in \cite{PhysRevE.87.042131}. Eq.~(\ref{performance}), which is the first main result of this paper, establishes then a quantum-thermodynamic limitation which holds for all the models of quantum absorption fridges existing in current literature, and is manifestly independent of the details of the working material of the refrigerators. While full details are deferred to the Methods section, we sketch a proof of the bound in what follows.

One may generically characterise the dissipation into a $d$-dimensional free bosonic field $\alpha$ at thermal equilibrium with flat spectral density, by decay rates of the form \cite{breuer2002theory}
\begin{equation}
\Gamma_{\omega,\alpha} =
\left\lbrace
	\begin{array}{ll}
		\gamma_0\omega^{d_\alpha}(1+N_\alpha(\omega))  & \omega > 0 \\
		\gamma_0\vert\omega\vert^{d_\alpha} N_\alpha(\vert\omega\vert) & \omega < 0
	\end{array}
\right.,
\label{spectral_correlation_tensor}
\end{equation}
where $N_\alpha(\omega)=[\exp{(\hbar\omega/k_B T_\alpha)}-1]^{-1}$. Eq.~\eqref{spectral_correlation_tensor} follows just from the application of the Born and Markov approximations to a general microscopic model of relaxation into a bosonic bath. From it, one can see that the cooling power of a three-level fridge [Fig.~\ref{Fig1}(a)] writes as
\begin{equation}
\dot{\mathcal{Q}}_c(x)=x^{d_c} p(x)\left(a_1 e^{-b_1 x} - a_2 e^{-b_2 x}\right),
\label{proof_sketch_0}\end{equation}
where $x\equiv\omega_c/\omega_{c,\max}$, $a_i$ and $b_i$ are positive real constants, and $p(x)$ is a positive function. The cooling power vanishes at both edges of the cooling window $\dot{\mathcal{Q}}_c(0)=\dot{\mathcal{Q}}_c(1)=0$ and is a concave function of $x$. Eq.~\eqref{proof_sketch_0} can always be conveniently recast as
\begin{equation}
\dot{\mathcal{Q}}_c(x)=P(x)~ x^{d_c} (1-x)\equiv P(x)f(x),
\label{proof_sketch_1}\end{equation}
with $P(x)$ again a positive function (see Eq.~\eqref{cold_heat_current4} in Methods). The maximum of $\dot{\mathcal{Q}}_c(x)$ is attained at some $x_*$ such that
\begin{equation}
\dot{\mathcal{Q}}_c'(x_*)=P'(x_*)f(x_*)+P(x_*)f'(x_*)=0,
\label{proof_sketch_2}\end{equation}
where the prime denotes differentiation with respect to $x$. Note that the positive and concave function $f(x)$ is maximised precisely at $x=d_c/(d_c+1)$.

\begin{figure*}[t]
	\subfigure{\label{Fig3a}
	\includegraphics[width=0.32\textwidth]{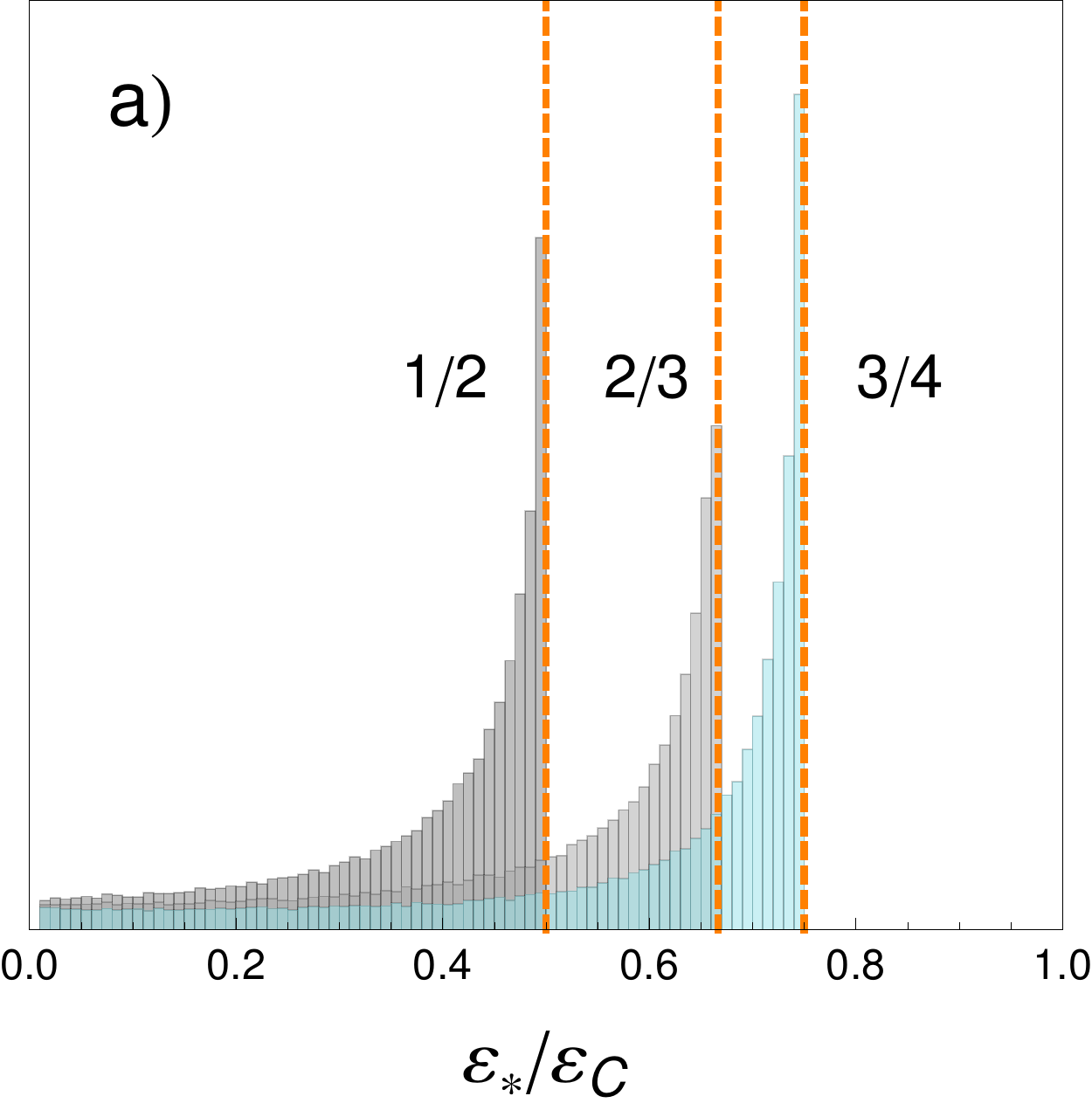}}
	\subfigure{\label{Fig3b}
	\includegraphics[width=0.32\textwidth]{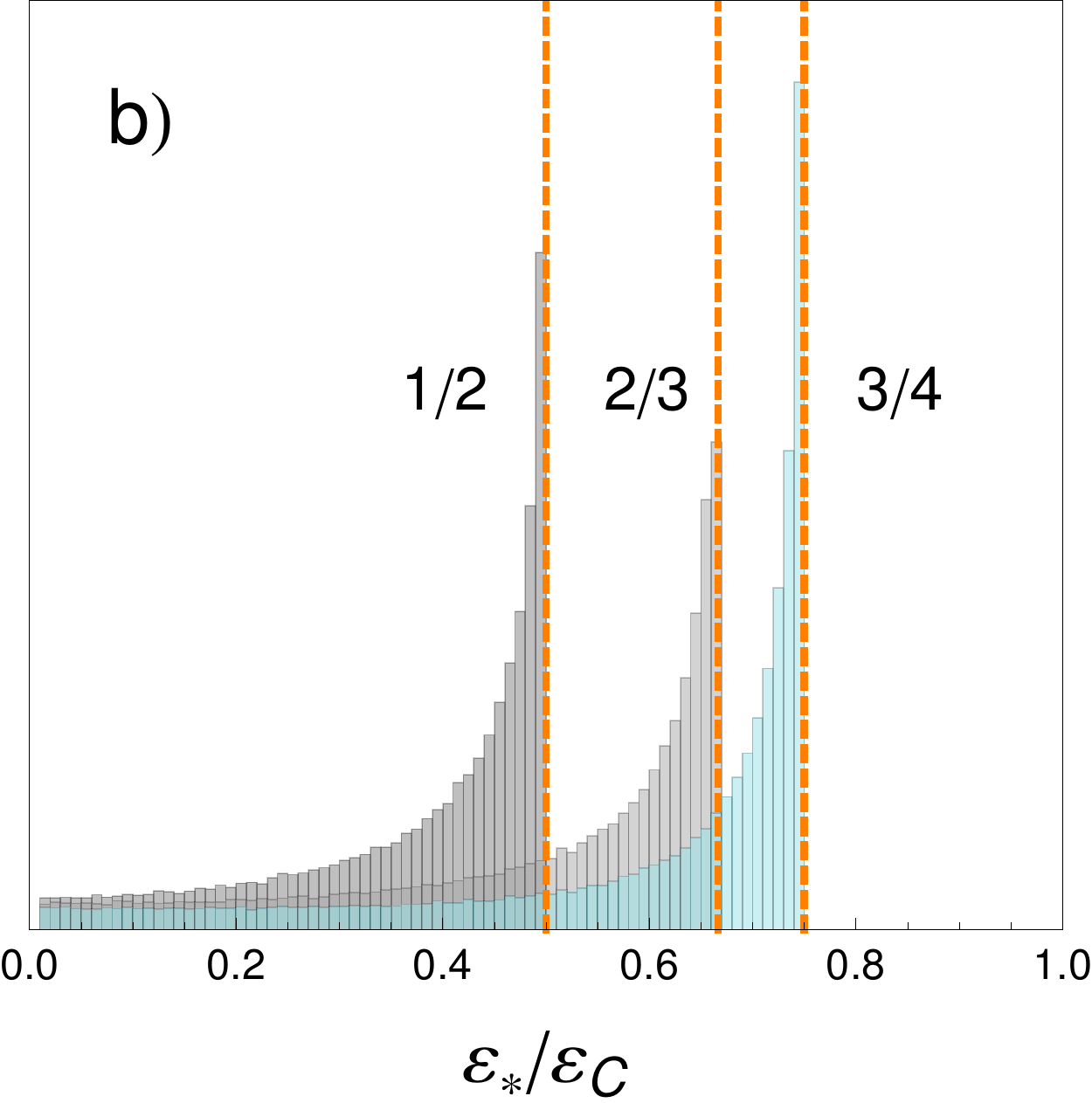}}
	\subfigure{\label{Fig3c}
	\includegraphics[width=0.32\textwidth]{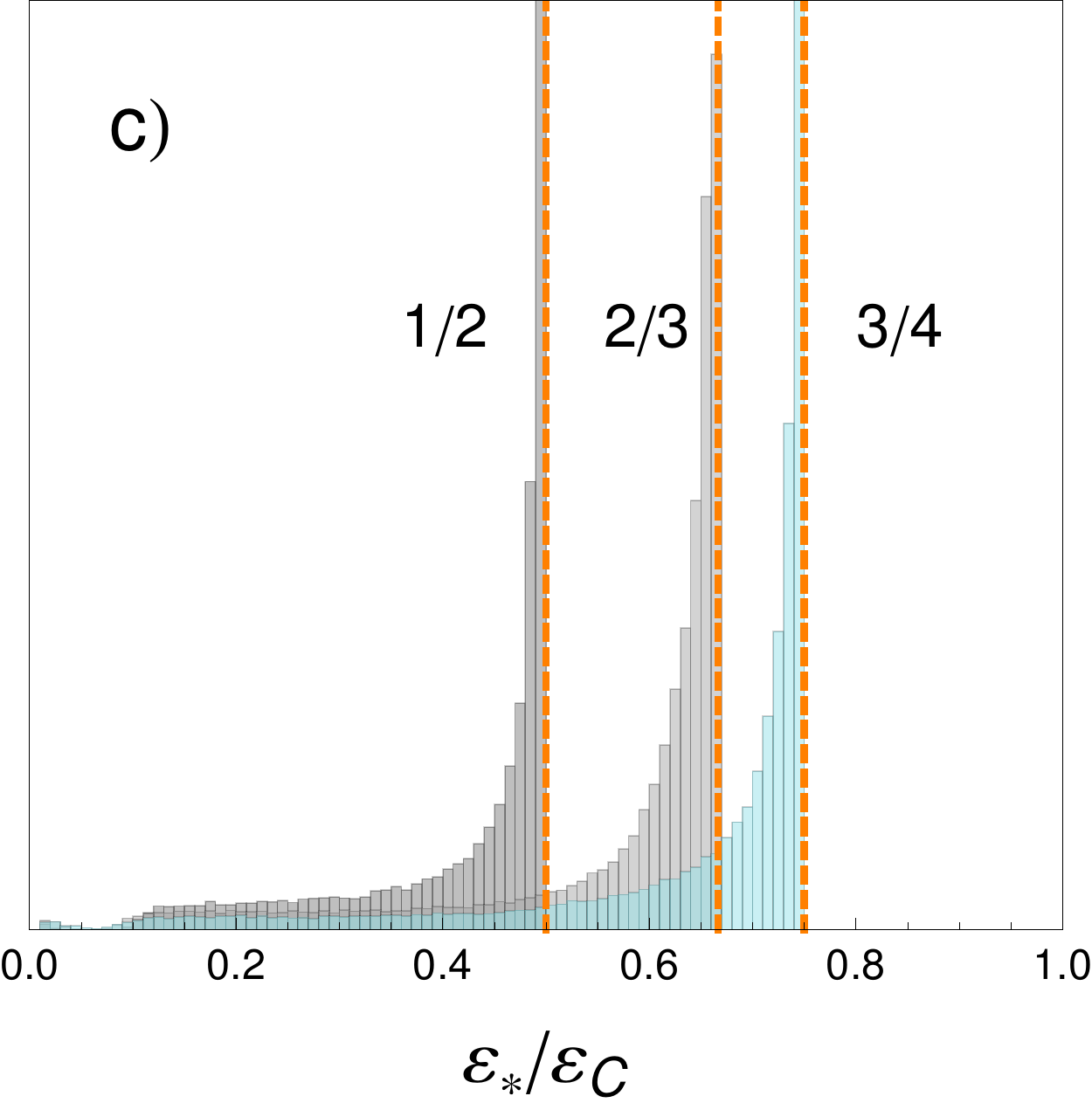}}
\caption{Histograms of the efficiency at maximum power $\varepsilon_*/\varepsilon_C$ for $\sim 10^5$ random (a) three-level, (b) two-qubit and (c) three-qubit absorption fridges, as introduced in Figs.~\ref{Fig1}(a), \ref{Fig1}(b), and \ref{Fig1}(c), respectively. The temperatures $\{T_\alpha\}$, the hot frequency $\omega_h$, the dissipation rates, and----for the case (c)---the interaction strength $g$,  were all chosen completely at random, but always respecting the conditions $\gamma\ll k_B T_\alpha$ and $\gamma\ll\{\hbar\omega_\alpha,g\}$, that guarantee the applicability of the master equation Eq.~\eqref{lindblad}. In each case, we found the optimal pair $\{\omega_w,\omega_c\}$ so that $\dot{\mathcal{Q}}_c$ was maximised, and computed the corresponding efficiency. This is equivalent to a global numerical optimisation of $\varepsilon_*$ over all free parameters of the models. In each of the panels above, we show three data sets that correspond to the same heat pump operating between $1d$, $2d$ and $3d$ baths. In agreement with Eq.~\eqref{performance}, $\varepsilon_*$ is bounded by $\frac12\varepsilon_C$, $\frac23\varepsilon_C$ and $\frac34\varepsilon_C$, respectively. It can also be seen that these bounds are tight, and that the majority of randomly sampled fridges cool very close to their corresponding bounds.}
\label{Fig3}
\end{figure*}

Previous literature on performance bounds for heat engines has established that the efficiency at maximum power is maximised in the limit of small $\varepsilon_C$ \cite{esposito2009universality,PhysRevLett.95.190602}, as it is also the case for quantum absorption fridges \cite{PhysRevE.87.042131}. We shall therefore expand $P(x)$ around $\varepsilon_C\rightarrow 0$, where in fact $P(x)\sim p(x)$ [see Eq.~\eqref{cold_heat_current4}], and study its analytical properties.
We then prove the claim by {\it reductio ad absurdum}. Let us first assume that $x_*>d_c/(d_c+1)$, so that the bound in Eq.~\eqref{performance} would not hold [recall Eqs.~\eqref{ideal} and \eqref{cooling_window}]. $P'(x)$ has at most one root in the interval $x\in[0,1]$ and can be seen to be negative in the neighbourhood of $x=d_c/(d_c+1)$, for vanishing $\varepsilon_C$. Being both $P(x)$ and $f(x)$ positive in the whole unit interval, the violation of the bound Eq.~\eqref{performance} would contradict Eq.~\eqref{proof_sketch_2}. Therefore, one must have instead $x_*\leq d_c/(d_c+1)$ and hence,
\begin{equation}
\varepsilon_*=\frac{\omega_{c,*}}{\omega_w}\leq\frac{d_c}{d_c+1}~\varepsilon_C,
\label{proof_sketch_3}\end{equation}
the bound being saturated if $P'(x_*)=0$.

Now, we shall consider the two-qubit model [Fig.~\ref{Fig1}(b)]. We find that its steady-state heat currents can be broken up as (see Methods for details)
\begin{subequations}
\begin{align}
\dot{\mathcal{Q}}_c &=\omega_c\left(q_1+q_2\right) \label{components_c}\\
\dot{\mathcal{Q}}_h &=-\omega_h\left(q_1+q_2\right) \label{components_h}\\
\dot{\mathcal{Q}}_w &=\omega_w\left(q_1+q_2\right) \label{components_w},
\end{align}
\label{components}\end{subequations}
where all $q_i>0$. There are two contributions to the total heat fluxes, namely $\vert\dot{\mathcal{Q}}_\alpha^{(1)}\vert=\omega_\alpha q_1$ and $\vert\dot{\mathcal{Q}}_\alpha^{(2)}\vert=\omega_\alpha q_2$, which individually satisfy the first and second law of Eqs.~\eqref{first_law} and \eqref{second_law} within the cooling window delimited by Eq.~\eqref{cooling_window}. Furthermore, $\vert\dot{\mathcal{Q}}_\alpha^{(i)}/\dot{\mathcal{Q}}_\beta^{(i)}\vert=\omega_\alpha/\omega_\beta$ for $i=\{1,2\}$ and hence, each three-level component behaves as an ideal refrigerator on its own.
Most importantly, the cold fluxes $\dot{\mathcal{Q}}_c^{(i)}$ have the same analytic structure of Eq.~\eqref{proof_sketch_0}, so that their efficiency at maximum power is limited by $\varepsilon_*^{(i)}/\varepsilon_C\leq d_c/(d_c+1)$. As a consequence, the performance of the combination of the two is also bounded by $\varepsilon_C~d_c/(d_c+1)$, since
\begin{equation}
\varepsilon_*=\frac{\varepsilon^{(1)}(x_*)\dot{\mathcal{Q}}_w^{(1)}(x_*)+\varepsilon^{(2)}(x_*)\dot{\mathcal{Q}}_w^{(2)}(x_*)}{\dot{\mathcal{Q}}_w^{(1)}(x_*)+\dot{\mathcal{Q}}_w^{(2)}(x_*)}\leq\frac{d_c}{d_c+1}\varepsilon_C .
\label{components_2}\end{equation}
The heat currents of the three-qubit fridge [Fig.~\ref{Fig1}(c)] are much more involved analytically. However, extensive numerical evidence confirms that Eq.~\eqref{performance} also applies to this case \cite{PhysRevE.87.042131} [see Fig.~\ref{Fig3}(c)].

A numerical investigation, presented in Fig.~\ref{Fig3}, certifies that the bound of Eq.~\eqref{performance} is {\it tight} for all the three models. One can further see that choosing $\omega_w/T_{w,h}\ll 1$ is a sufficient condition to approach it closely under large temperature differences $T_c/T_h \ll 1$. These are to be regarded as analytical design prescriptions for the practical implementation of optimal quantum heat pumps, e.g.~following recent proposals involving  superconducting qubits or arrays of quantum dots \cite{0295-5075_97_4_40003,PhysRevLett.110.256801}. Note as well that the limit $T_c/T_h \ll 1$ also implies $\varepsilon_*<\varepsilon_C \ll 1$, as should be expected.

Our bound has been analytically established as a constraint on the performance of the fridge operation mode of the three-level maser and that of any ideal absorption refrigeration cycle reducible to three-level systems, and it has been also shown to hold for fundamentally different non-ideal devices. It is in order to remark, however, that  Eq.~(\ref{performance}) has in fact a \textit{general} validity transcending specific models. Let us consider a totally generic model of quantum absorption fridge: We can always write the leading contribution to its cooling power $\dot{\mathcal{Q}}_c$ as a sum of terms of the form
\begin{equation}
\dot{\mathcal{Q}}_c \sim \hbar\omega_c\left(\mathscr{P}_{\omega_c}^{\uparrow}\mathscr{P}_{\omega_w}^{\uparrow}\mathscr{P}_{\omega_h}^{\downarrow}-
\mathscr{P}_{\omega_c}^{\downarrow}\mathscr{P}_{\omega_w}^{\downarrow}\mathscr{P}_{\omega_h}^{\uparrow}\right),
\label{general_power}\end{equation}
where $\mathscr{P}^{\uparrow\downarrow}_{\omega_\alpha}$ stands for excitation/relaxation rates of the contact transitions at $\omega_\alpha$. Eq.~\eqref{general_power} just formalises the imbalance between the elementary three-stroke cooling and heating cycles which has to underly any implementation of quantum absorption cooling, and contains no details on the specific working material of the refrigerator, nor about the spectral properties of the reservoirs to which it couples. Given that these three reservoirs are in thermal equilibrium, we can make use of the detailed balance condition of Eq.~\eqref{detailed_balance} and thus arrive to
\begin{equation}
\dot{\mathcal{Q}}_c \sim \hbar\omega_c~\prod_\alpha\mathscr{P}_{\omega_\alpha}^\downarrow\left(e^{-\frac{\hbar\omega_c}{k_B T_c}}e^{-\frac{\hbar\omega_w}{k_B T_w}}-e^{-\frac{\hbar\omega_h}{k_B T_h}}\right),
\label{general_power2}\end{equation}
which is formally identical to Eq.~\eqref{proof_sketch_0}, with the positive function $p(x)\propto \prod_\alpha\mathscr{P}_{\omega_\alpha}^\downarrow(x)/x^{d_c-1}$. Therefore, the whole line of reasoning between Eqs.~\eqref{proof_sketch_0}--\eqref{proof_sketch_3} is still applicable,  provided that the condition $p'(x_*)\leq 0$ is verified. This weak assumption is sufficient for Eq.~\eqref{performance} to hold, regardless of the physical support of the refrigerator and the specific properties of the baths. Under these premises, we conclude that our bound is \textit{model-independent} for quantum absorption refrigerators.

Let us eventually connect this result with the other known bounds from finite-time thermodynamics. In the seminal work by Curzon and Ahlborn \cite{curzon1975efficiency}, an upper limit to the efficiency at maximum power was derived for a Carnot engine under the \textit{endoreversible} approximation \cite{hoffmann1997endoreversible}. In spite of its seemingly limited scope, such bound succeeds in capturing the universal behaviour of $\varepsilon_*$ at small efficiencies \cite{esposito2009universality,PhysRevLett.95.190602}, and recurrently appears in different models of heat engines \cite{PhysRevLett.105.150603,wang2012finite,geva1991spin,esposito2009universality,esposito2010universalityCA}, including the three-level maser in its engine configuration \cite{esposito2009universality}. However, the obtention of similar results for refrigerators would require a more careful phenomenological modeling of their main sources of irreversibility, and can lead to highly model-dependent performance bounds \cite{Wu1996299,chen1994new,PhysRevA.39.4140}.

In analogy with the Curzon-Ahlborn limit, our result in Eq.~\eqref{performance} accurately represents $\varepsilon_*$ at low efficiencies for all embodiments of quantum absorption refrigerators, which encompass fundamentally different models. The same limit holds as a strict upper bound to the efficiency at maximum power of `classical' endoreversible absorption chillers \cite{PhysRevA.39.4140}. It must be noted as well, that it relies on a consistent microscopic description of the system-baths interactions rather than on a phenomenological ansatz. From the physical point of view, such a model-independent bound can be understood from the intuitive notion that the maximisation of the cooling power in the steady state must be governed by the low-frequency dependence of the corresponding cold decay rate [see Eq.~\eqref{spectral_correlation_tensor}].
\newline

\noindent\textbf{Superefficiency: Squeezing the second law.}
Up to now, we have seen how the standard laws of thermodynamics place fundamental constraints on heat pumps, even when they are made up of a single finite-dimensional open quantum system. This is perhaps not surprising taking into account that there is nothing really `quantum' about the operation of these devices, other than the discreteness of their energy spectra \cite{PhysRevE.87.042131}. For instance, the ideal fridges of Figs.~\ref{Fig1}(a) and \ref{Fig1}(b) operate in completely classical (diagonal) steady states, while the quantum coherence that builds up asymptotically in the non-ideal fridges of Fig.~\ref{Fig1}(c) does not seem to affect their performance in any crucial way. Even if steady-state bipartite entanglement may exist in this case, it usually appears only in the regime of very low efficiencies and vanishing cooling power \cite{1305.6009v1}. Likewise, other types of quantum correlations, though widely present, do not have any influence on optimal cooling \cite{PhysRevE.87.042131}.

Is it then possible at all for quantum heat pumps to operate past the `classical' limits? Ideally, one would like to devise tricks to push their performance bounds further, of course remaining always within the standard framework of quantum thermodynamics and not advocating any violation of its laws. We shall devote this section to illustrate how one can indeed go beyond Eqs.~\eqref{efficiency} and \eqref{performance}, by exploiting non-equilibrium environmental features that can be mimicked with suitable quantum reservoir engineering techniques.

The whole idea would consist in initialising the work reservoir in a \textit{squeezed-thermal} state
\begin{equation}
\hat \chi^{T,r}_w\equiv\hat S(\xi)~\hat \chi_w^{T}~\hat{S}^{\dagger}(\xi).
\label{squeezed_state}\end{equation}
Such a state results from the action of the unitary squeezing operator \cite{walls2008quantum}
\begin{equation}
\hat S(\xi)\equiv\prod_\mu \exp\left\lbrace\frac{1}{2}\xi^* \hat b_\mu^2-\frac{1}{2}\xi (\hat b^{\dagger}_\mu)^2 \right\rbrace
\label{squeezing}\end{equation}
on a thermal preparation $\hat \chi^{T}_w$,  where $\{\hat b_\mu,\hat b^{\dagger}_\mu\}$ stand for creation and annihilation operators on mode $\mu$ of the work reservoir. The parameter $\xi\equiv r\exp{i\theta}$ is in general a complex number, although for our purpose we can set the phase $\theta$ to zero and consider a real squeezing parameter   $\xi=r\in\mathbb{R}$. The \textit{quantumness} of a squeezed state resides in the asymmetry of the variances of its field quadratures $\hat X^\mu_\pm=\hat b_\mu\pm \hat b^{\dagger}_\mu$, as opposed to a (classical) coherent state, for which $\Delta\hat X_+^\mu = \Delta\hat X_-^\mu$.

It is well known that squeezed states, even in absence of entanglement, allow for quantum enhancement in several applications of information theory \cite{RevModPhys.58.1001}, including quantum cryptography \cite{PhysRevA.61.022309},  and most notably precision measurements \cite{PhysRevD.23.1693} and quantum metrology below shot noise \cite{giovannetti2011advances}, with applications e.g.~for gravitational wave interferometry \cite{PhysRevLett.104.251102,aasi2013enhanced,abadie2011gravitational}.

The key property to be exploited here is the non-stationarity of squeezed preparations. This results in a periodic time modulation of the correlation functions of the work reservoir, which is somewhat equivalent to the action of an external driving on the heat pump. In this sense, the work reservoir may now play an active role in the cooling process. As we shall now see, this is achieved without compromising the thermodynamic consistency of the whole setting.

In order to account for the squeezing of the work reservoir, the quantum master equation Eq.~\eqref{lindblad} must be replaced by
\begin{equation}
\frac{d}{dt}~\hat\varrho(t)=\left(\mathscr{L}_w^r+\mathcal{L}_h+\mathcal{L}_c\right)\hat{\varrho}(t),
\label{lindblad_sqz}\end{equation}
where the modified work dissipator $\mathscr{L}_w^r\equiv\mathcal{L}_w^r+\Delta\mathcal{L}_w^r$ can be cast in the standard LGKS form \cite{breuer2002theory} (see Methods for details).

Clearly, if the work heat current is suitably redefined as $\dot{\mathcal{Q}}_w=\text{tr}[\hat H_\text{wm}\mathscr{L}_w^r\hat \varrho(\infty)]$, the asymptotic stationarity of energy implies again the first law as stated in Eq.~\eqref{first_law}.

However, the dissipator $\mathscr{L}_w^r$ will not generally bring the corresponding contact transition to thermal equilibrium at temperature $T_w$, but to some other steady state $\hat\sigma_w(r)$. Still, the LGKS form of the overall generator of the dynamical map in Eq.~\eqref{lindblad_sqz} guarantees its full contractivity \cite{spohn1978entropy}, so that the second law generalises to
\begin{equation}
-\text{tr}\left[\mathscr{L}_w^r\hat \varrho(\infty)\log\hat\sigma_w(r)\right]+\frac{\dot{\mathcal{Q}}_h}{T_h}+\frac{\dot{\mathcal{Q}}_c}{T_c}\leq 0.
\label{second_law_sqz}\end{equation}
It is convenient to fit $\hat\sigma_w(r)$ by a thermal state of the form $\hat\sigma_w(r)\propto\exp\{-\hat H_\text{wm}/k_B T^\text{eff}_w(r)\}$, with a suitable squeezing-dependent work temperature $T^\text{eff}_w(r)$. Since the contact with the work reservoir only involves two levels, this is always possible. Eq.~\eqref{second_law_sqz} may be thus rewritten in the more familiar way
\begin{equation}
\frac{\dot{\mathcal{Q}}_w}{T_w^\text{eff}(r)}+\frac{\dot{\mathcal{Q}}_h}{T_h}+\frac{\dot{\mathcal{Q}}_c}{T_c}\leq 0.
\label{second_law_sqz2}\end{equation}
For any $r>0$, the effective temperature $T^\text{eff}_w(r)$ exceeds $T_w$, and it diverges for $r\rightarrow\infty$. Therefore, the generalised squeezed-dependent Carnot efficiency will always exceed its classical value
\begin{equation}
\varepsilon_C(r)=\frac{T_c}{T_h-T_c}\left(1-\frac{T_h}{T_w^\text{eff}(r)}\right)>\varepsilon_C(0).
\label{carnot_sqz}\end{equation}
Note that, as should be expected, in the limit of $r\rightarrow\infty$, Eq.~\eqref{carnot_sqz} saturates to the maximum efficiency of a power-driven refrigerator (i.e. a reversed heat engine) $\varepsilon_C(\infty)=T_c/(T_h-T_c)$ \cite{PhysRev.156.343}.

\begin{figure}[t]
	\includegraphics[width=0.9\columnwidth]{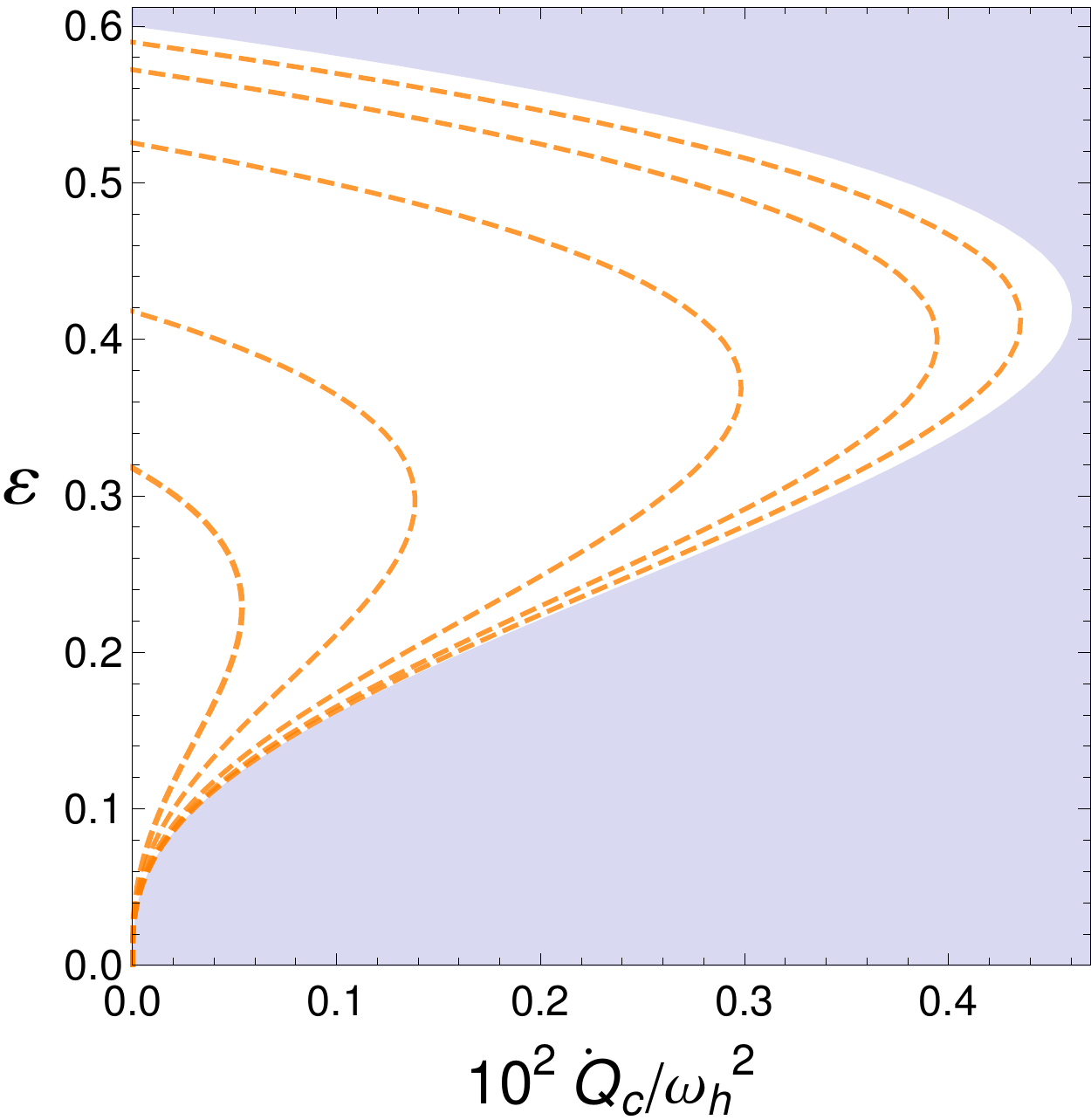}
	\caption{Parametric plot of power vs. efficiency of a three-level maser operating between three-dimensional bosonic reservoirs at temperatures $T_w=170$, $T_h=80$ and $T_c=30$ ($\hbar=k_B=1$), for different values of the squeezing parameter of the work reservoir $r=\{0,\,0.5,\,1,\,1.5,\,2\}$. Both $\dot{\mathcal{Q}}_c$, $\varepsilon_*$ and the maximum $\varepsilon$ increase with $r$. In each curve, the hot frequency was fixed at $\omega_h=50$, while $\omega_c$ ranges from $0$ up to $\omega_{c,\max}(r)$. The frontier between the white and the shaded region corresponds to the performance characteristic of an ideal reversed heat engine operating between the same hot and a cold baths. The general behaviour illustrated here is totally independent of the specific values of $\omega_h$ and the set of temperatures $\{T_\alpha\}$.}
	\label{Fig4}
\end{figure}

Hence, we have demonstrated how, for a given set of thermal resources $\{T_c,T_h,T_w\}$, classically forbidden superefficient quantum absorption cooling may be realised by just squeezing the work reservoir. This is illustrated in Fig.~\ref{Fig4}, with a set of performance characteristics corresponding to different environmental squeezing parameters in the three-level maser heat pump of Fig.~\ref{Fig1}(a). Perhaps even more remarkable than the seemingly counter-intuitive possibility of cooling above the classical efficiency threshold, is the fact that the efficiency at maximum power {\it and} the cooling power itself \emph{systematically increase} jointly with the squeezing parameter. This is a striking consequence of the non-equilibrium environmental fluctuations.

In the macroscopic domain, the practical drawbacks of an absorption chiller \cite{einstein1927refrigeration}, as compared to an equivalent work-driven refrigeration cycle, are typically its much lower efficiency and output power. Nevertheless, as we can see in Fig.~\ref{Fig4}, a quantum-enhanced heat-driven fridge does not only outperform its purely thermal counterpart, but is also capable of approaching very closely the performance of a Carnot fridge, when provided with a \emph{finite} amount of environmental squeezing. Thus, as opposed to the classical case, quantum absorption cooling may compete on an equal footing with conventional refrigeration, when combined with reservoir engineering techniques \cite{PhysRevLett.77.4728,PhysRevA.57.548,shahmoon2013engineering} to allow for the exploitation of non-equilibrium squeezed environmental fluctuations. This observation could be a promising first step towards a number of applications of heat-driven  refrigeration to a new generation of quantum and nanoscale technologies.

\section*{Discussion}
Let us start by summarising the two main results of this paper. On the one hand, we proved that the efficiency at maximum power of all the known models of quantum absorption refrigerator (both ideal and non-ideal) is tightly upper bounded by a fraction of the Carnot efficiency $\varepsilon_C$, which is independent of the details of the device and only relates to the spectral properties of the thermal fluctuations of the environment. On the other hand, we showed most remarkably how by squeezing the heat source that drives an absorption cooling cycle, one may boost its performance to the extent of making it comparable to a conventional power-driven cooling device.

A key path to the optimisation of autonomous quantum heat pumps is thus found to reside in applying suitable reservoir engineering techniques \cite{PhysRevLett.77.4728,PhysRevA.57.548,shahmoon2013engineering} rather than in exploiting any resource intrinsic to their quantum mechanical working materials \cite{PhysRevE.87.042131,1305.6009v1}. Indeed, the importance of reservoir manipulation has been very recently acknowledged in the context of quantum thermodynamics: It has been recently speculated that superefficient operation of quantum heat engines, in apparent violation of the second law, may be achieved e.g.~by reservoir squeezing \cite{PhysRevE.86.051105} or using more general types of non-equilibrium reservoirs \cite{dillenschneider2009energetics,1303.6558v1}, and by connecting the working material to an adiabatically isolated ancilla \cite{PhysRevA.87.063845}. However, to our knowledge, this paper represents the first demonstration of a \textit{systematic} enhancement in the performance of a quantum-mechanical thermal device. Furthermore, this unconditional enhancement is, in principle, achievable without manipulating the working material or the given hot and cold baths: It is only the arguably controllable heat source that needs to be tailored.

At this point, one could wonder if generating squeezing in the work reservoir is really worth the effort when one could raise instead its equilibrium temperature to obtain effectively the same results. Back to Fig.~\ref{Fig4}, we see that a squeezing parameter of $r=1.5$ ($\sim 13$ dB, which is currently at reach \cite{PhysRevLett.104.251102}), would already take the heat pump close to its best equivalent power-driven counterpart. In order to achieve a comparable amplification without squeezing, one should increase the work temperature by at least one order of magnitude, which might just not be possible in engineered or natural environments. Thus, it seems reasonable that in many concrete situations, reservoir squeezing could be indeed the best and most natural way to boost the performance of a heat-driven fridge. However, supporting the claim that a quantum-enhanced absorption refrigerator can really compete with a heat pump driven by work is a more delicate issue: To mimic a non-stationary state for the work bath \cite{PhysRevLett.77.4728,PhysRevA.57.548,shahmoon2013engineering}, one has pay an extra cost that should be added to the overall efficiency balance for a fair comparison. Again, depending of the specifics of the implementation, it may well be that a quantum-enhanced quantum absorption chiller came to cool \textit{cheaper} than the corresponding quantum `compression' cycle. Furthering this issue demands a study on its own.

We also wish to make clear the realm of validity of our analysis. In all of the above, we have taken the fulfilment of the first and second laws as a guarantee of thermodynamic consistency, but we have neither discussed the third law, nor probed the neighbourhood of the absolute zero \cite{PhysRevE.85.061126}. Since we approach the open system dynamics from a LGKS-type master equation, the Markov approximation should better hold, which in turn implies that the thermal fluctuations must be sufficiently fast as compared with the dissipation time scales. Thus, to be always on the safe side, we must limit our environmental temperatures from below.

When it comes to the experimental realisation of quantum-enhanced absorption technologies it is worth mentioning the pioneering refrigeration experiments by Geusic et al.~\cite{5123190} using the energy level structure of Cr$^{3+}$ ions in a ruby crystal as support for a three-level maser. As already mentioned, detailed proposals exist as well for the non-ideal three-qubit model of Fig.~\ref{Fig1}(c) using superconducting qubits \cite{0295-5075_97_4_40003}, and quantum dots \cite{PhysRevLett.110.256801}. The squeezing can be engineered by coherently driving the work transition, potentially involving auxiliary levels and vacua; feasible schemes have been proposed e.g.~involving trapped ions or Rydberg atoms \cite{PhysRevLett.77.4728,PhysRevA.57.548,shahmoon2013engineering}.

In conclusion, we have demonstrated the possibility of genuine quantum-enhanced absorption refrigeration beyond the fundamental bounds imposed by classical thermal environments. The application of reservoir engineering techniques to autonomous quantum heat pumps might render them practically useful and competitive for many applications of quantum technologies, {\it in primis} quantum cooling. Moreover, the distinctive simplicity of these devices is ideal to get a clean glimpse of how thermodynamics looks like beyond the standard scenario.

\section*{Methods}

\noindent\textbf{Quantum master equation.} We will now write down explicitly the equation of motion for the working material of a quantum heat pump, introduced in Eqs.~\eqref{lindblad} and \eqref{lindblad_sqz}. The total Hamiltonian is generically (see caption of Fig.~\ref{Fig1})
\begin{equation}
\hat H = \hat H_\text{wm} + \sum_\alpha \hat H_{B_\alpha} + \sum_\alpha \hat H^\text{int}_\alpha,
\label{hamiltonian}\end{equation}
where the dissipative system-reservoir interactions $\hat H^\text{int}_\alpha$ may be taken as
\begin{equation}
\hat H^\text{int}_\alpha\propto\left( \meket{0}{\alpha}\mebra{1}{\alpha} + \meket{1}{\alpha}\mebra{0}{\alpha} \right) \otimes \hat{\mathcal{B}}_\alpha\equiv\hat\sigma_\alpha\otimes\hat{\mathcal{B}}_\alpha.
\label{interaction}\end{equation}
Here, the ground and excited states of the two-level \textit{contact port} with bath $\alpha$ are denoted by $\meket{0}{\alpha}$ and $\meket{1}{\alpha}$ respectively. Let us first assume that all three reservoirs are prepared in thermal equilibrium. Under the further assumptions of vanishing initial correlations between system and environments, weak system-environment interaction and separation of time scales of free and dissipative dynamics (Born, Markov and rotating-wave approximations), one arrives to the well known LGKS-type quantum master equation \cite{breuer2002theory} with dissipators as those of Eq.~\eqref{Lindblad_dissipator}.

The elements of the spectral correlation tensor, given explicitly in Eq.~\eqref{spectral_correlation_tensor}, follow from the power spectrum of the environmental correlations $\Gamma_{\omega,\alpha}=2~\text{Re}\int_0^\infty ds~e^{i\omega s}\langle\hat{\mathcal{B}}^\dagger_\alpha(t)\hat{\mathcal{B}}_\alpha(t-s)\rangle$. On the other hand, $\hat A_\alpha(\omega)$ result from the decomposition of $\hat\sigma_\alpha$ as eigenoperators of $\hat H_\text{wm}$ \cite{breuer2002theory}, and the discrete index $\omega$ labels all the open decay channels, i.e. all the energy differences that correspond to non-vanishing jump operators.

In the case of an ideal fridge, for which $\meket{0}{\alpha}$ and $\meket{1}{\alpha}$ are picked among the eigenstates of $\hat H_\text{wm}$, Eq.~\eqref{lindblad} rewrites as
\begin{multline}
\frac{d\hat\varrho}{dt} = \sum_\alpha\sum_{\omega\in\Omega} \Gamma_{\alpha,\omega} \left(\hat\sigma^{-}_\alpha~\hat\varrho~\hat\sigma^{+}_\alpha - \frac{1}{2} \left\lbrace \hat \sigma^{+}_\alpha\hat \sigma^{-}_\alpha,\hat \varrho \right\rbrace_+ + \right. \\
\left. e^{-\hbar\omega/k_B T_\alpha}\hat\sigma^{+}_\alpha~\hat\varrho~\hat\sigma^{-}_\alpha - \frac{1}{2}e^{-\hbar\omega/k_B T_\alpha} \left\lbrace \hat \sigma^{-}_\alpha\hat \sigma^{+}_\alpha,\hat \varrho \right\rbrace_+\right),
\label{lindblad2}\end{multline}
where $\hat\sigma^{+}_\alpha=\meket{1}{\alpha}\mebra{0}{\alpha}$ and $\hat\sigma^{-}=(\hat\sigma^+)^\dagger$ and the inner summation runs over the frequencies $\Omega=\{\omega_w,\omega_h,\omega_c\}$. Eq.~\eqref{lindblad2} accounts for the reduced dynamics of the three-level and two-qubit heat pumps of Figs.~\ref{Fig1}(a) and \ref{Fig1}(b). The case of the non-ideal three-qubit model of Fig.~\ref{Fig1}(c) is more involved. All the details may be found elsewhere \cite{PhysRevE.87.042131}.

Let us now relax the assumption of equilibrium environments to allow for squeezing in the work reservoir. The key difference is that the rates $\Gamma_{\alpha,\omega}$ become explicitly time-dependent as a consequence of the non-stationarity of squeezed preparations. After performing the rotating-wave approximation, new terms appear in the work dissipator \cite{breuer2002theory}, that becomes
\begin{multline}
\mathscr{L}_w^r=\mathcal{L}^r_w+\Delta\mathcal{L}^r_w = \\ \sum_{\omega} \Gamma_{w,\omega}^{r} \left(\hat A_w(\omega) \hat \varrho \hat A_w^\dagger(\omega) - \frac{1}{2} \left\lbrace \hat A^\dagger_w(\omega)\hat A_w(\omega),\hat \varrho \right\rbrace_+\right) + \\
+ \Lambda_{w,\omega}^{r} \left(\hat A_w(\omega) \hat \varrho \hat A_w(\omega) + \hat A^\dagger_w(\omega) \hat \varrho \hat A^\dagger_w(\omega)\right.  \\  \left. - {\hat A_w(\omega)}^2\hat\varrho- \hat\varrho{\hat A^\dagger_w(\omega)}^2\right),
\label{lindblad_sqz2}\end{multline}
where the squeezing-dependent coefficients $\Gamma_{w,\omega}^r$ and $\Lambda_{w,\omega}^r$ are given by
\begin{subequations}
\begin{equation}
\Gamma_{w,\omega}^r =
\left\lbrace
	\begin{array}{ll}
		\gamma_0\omega^{d_w} (N_w^r(\omega)+1)  & ~~\omega > 0 \\
		\gamma_0\vert\omega\vert^{d_w} N_w^r(\vert\omega\vert)  & ~~\omega < 0
	\end{array}
\right.
\label{spectralct_sqz1}\end{equation}
\begin{equation}
\Lambda_{w,\omega}^r = \frac{1}{2}\gamma_0\vert\omega\vert^{d_w} M_w^r(\vert\omega\vert),
\label{spectralct_sqz2}\end{equation}
\end{subequations}
with $N_w^r(\omega)=N_w(\omega)\cosh{2r}+\sinh{r}^2$ and $M_w^r=\sinh{2r}~(2N_w(\omega)+1)/2$. A new set of jump operators can always be found so that the work dissipator in Eq.~\eqref{lindblad_sqz2} takes the standard LGKS form of Eq.~\eqref{Lindblad_dissipator}, which in turn, guarantees that the effective dynamics of the working material remains that of a \textit{dynamical semigroup} (i.e. completely positive and trace-preserving) in spite of the squeezing \cite{lindblad1976generators}.
\newline

\noindent\textbf{The second law at steady state.} It is known that the entropy production of a dynamical semigroup is a strictly positive quantity \cite{spohn1978entropy}. It is defined as
\begin{equation}
\Sigma\left[\hat\varrho(t)\right]\equiv-\frac{d}{dt} S\left(\hat\varrho(t)\vert\vert\hat\varrho^0\right)\geq 0,
\label{entropy_production}\end{equation}
where $S(\hat\varrho\vert\vert\hat\varrho^0)=\text{tr}[\hat\varrho(\log\hat\varrho-\log\hat\varrho^0)]$ is the quantum relative entropy and $\hat\varrho^0$ is a steady state of the dissipative dynamics. Each of the dissipators appearing in Eq.~\eqref{lindblad} individually satisfies the inequality
\begin{equation}
\text{tr}\left[\left(\mathcal{L}_\alpha\hat\varrho\right)\log\hat\varrho_\alpha^0\right]-\text{tr}\left[\left(\mathcal{L}_\alpha\hat\varrho\right)\log\hat\varrho\right]\geq 0,
\label{second_law_der1}\end{equation}
where $\hat\varrho^0_\alpha$ is now stationary to the \emph{local} dissipator $\mathcal{L}_\alpha$ alone, i.e. $\mathcal{L}_\alpha\hat\varrho_\alpha^0=0$. Now, summing Eq.~\eqref{second_law_der1} over $\alpha$ yields
\begin{equation}
\sum_\alpha\text{tr}\left[\left(\mathcal{L}_\alpha\hat\varrho\right)\log\hat\varrho_\alpha^0\right]+\frac{d}{dt}S(\hat\varrho)\geq 0,
\label{second_law_der2}\end{equation}
where $S(\hat\varrho)=-\text{tr}[\hat\varrho\log\hat\varrho]$ stands for the von Neumann entropy. We may replace $\hat\varrho$ in Eq.~\eqref{second_law_der2} with the steady state of the full dynamics $\hat\varrho(\infty)$, and the local steady states $\hat\varrho_\alpha^0$, with equilibrium states $e^{-\hat H_\text{wm}/k_B T_\alpha}$. Note that indeed $\mathcal{L}_\alpha e^{- \hat H_\text{wm}/k_B T_\alpha}=0$ \cite{spohn1978entropy}. We thus obtain
\begin{equation}
-\sum_\alpha\frac{\text{tr}[\hat H_\text{wm}\mathcal{L}_\alpha\hat\varrho(\infty)]}{k_B T_\alpha}\geq 0\Rightarrow\frac{\dot{\mathcal{Q}}_w}{T_w}+\frac{\dot{\mathcal{Q}}_h}{T_h}+\frac{\dot{\mathcal{Q}}_c}{T_c}\leq 0,	
\label{second_law_der3}\end{equation}
that is, we recover the second law as stated in Eq.~\eqref{second_law}. When the work reservoir is prepared in a squeezed thermal state, the derivation remains exactly the same, only replacing $T_w$ with the squeezing-dependent effective temperature \cite{PhysRevE.86.051105}
\begin{equation}
T^\text{eff}_w(r)=\frac{\hbar\omega_w}{k_B\log\frac{\tanh^2{r}+\exp{\hbar\omega_w/k_B T_w}}{1+\tanh^2{r}\exp{\hbar\omega_w/k_B T_w}}}\geq T_w.
\label{effective_temperature}\end{equation}
$T_w^\text{eff}(r)$ is such that $\mathscr{L}^r_w e^{-\hat H_\text{wm}/k_B T_w^\text{eff}(r)}=0$. This leads to the modified second law of Eq.~\eqref{second_law_sqz}.
\newline

\noindent\textbf{Efficiency at maximum power of a three-level fridge.} We will devote this section to prove the ultimate bound $\varepsilon_*\leq d_c/(d_c+1)\varepsilon_C$ on the efficiency at maximum power for the three-level prototype of absorption refrigerator of Fig.~\ref{Fig1}(a). From Eq.~\eqref{lindblad2}, the steady state of the three level system may be readily found to be
\begin{subequations}
\begin{align}
\varrho_{11}(\infty) &= \frac{\Gamma_{\omega_c}\Gamma_{\omega_w}+\Gamma_{\omega_c}\Gamma_{\omega_h}+\Gamma_{-\omega_w}\Gamma_{\omega_h}}{\Delta} \label{varrho11}\\
\varrho_{22}(\infty) &= \frac{\Gamma_{-\omega_h}\Gamma_{\omega_w}+\Gamma_{-\omega_c}\Gamma_{\omega_w}+\Gamma_{-\omega_c}\Gamma_{\omega_h}}{\Delta} \label{varrho22}\\
\varrho_{33}(\infty) &= \frac{\Gamma_{-\omega_h}\Gamma_{\omega_c}+\Gamma_{-\omega_h}\Gamma_{-\omega_w}+\Gamma_{-\omega_c}\Gamma_{\omega_w}}{\Delta} \label{varrho33}.
\end{align}

Here, the ground, first and second excited states of $\hat H_\text{wm}$ are labeled $\ket{1}$, $\ket{2}$ and $\ket{3}$ respectively and all coherences asymptotically vanish. $\Gamma_{\omega_\alpha}$ is a shorthand for $\Gamma_{\alpha,\omega_\alpha}$ and the denominator $\Delta$ is given by
\begin{multline}
\Delta = \Gamma_{\omega_c}\Gamma_{\omega_w}+\Gamma_{-\omega_h}\Gamma_{\omega_w}+\Gamma_{-\omega_c}\Gamma_{\omega_w} \\
+\Gamma_{\omega_c}\Gamma_{\omega_h}+\Gamma_{-\omega_w}\Gamma_{\omega_h}+\Gamma_{-\omega_c}\Gamma_{\omega_h} \\
+\Gamma_{-\omega_h}\Gamma_{\omega_c}+\Gamma_{-\omega_h}\Gamma_{-\omega_w}+\Gamma_{-\omega_c}\Gamma_{-\omega_w}.
\label{steady_state_qutrit2}\end{multline}
\label{steady_state_qutrit}\end{subequations}
Combining the excitation and relaxation rates $\Gamma_{-\omega_c},\Gamma_{\omega_c}$ of the cold transition with its steady-state populations $\varrho_{11}(\infty),\varrho_{22}(\infty)$ one obtains a stationary heat current $\dot{\mathcal{Q}}_c$ given by
\begin{multline}
\dot{\mathcal{Q}}_c=\omega_c\left(\Gamma_{-\omega_c}\varrho_{11}(\infty)-\Gamma_{\omega_c}\varrho_{22}(\infty)\right)\\
=\omega_c\frac{\Gamma_{\omega_h}\Gamma_{-\omega_w}\Gamma_{-\omega_c}-\Gamma_{-\omega_h}\Gamma_{\omega_w}\Gamma_{\omega_c}}{\Delta},
\label{cold_heat_current}\end{multline}
which, using the detailed balance relations, becomes (from now on we will set $\hbar=k_B=1$)
\begin{equation}
\dot{\mathcal{Q}}_c=\omega_c\Gamma_{\omega_h}\Gamma_{\omega_w}\Gamma_{\omega_c}\frac{e^{-\omega_c/T_c}e^{-\omega_w/T_w}-e^{-\omega_h/T_h}}{\Delta}.
\label{cold_heat_current2}\end{equation}
Given the constraint $\omega_h=\omega_c+\omega_w$, we may fix the values of $\{\omega_w,T_w,T_h,T_c\}$ and take $x\equiv\omega_c/(\omega_w\varepsilon_C)$ as the only independent variable in Eq.~\eqref{cold_heat_current2}. Of course, one would obtain the same results by fixing $\omega_h$ instead of $\omega_w$. This brings us back to Eq.~\eqref{proof_sketch_0}
\begin{equation}
\dot{\mathcal{Q}}_c(x)=x^{d_c} p(x)\left(a_1 e^{-b_1 x} - a_2 e^{-b_2 x}\right),
\label{cold_heat_current3}\end{equation}
where $a_1=e^{-\omega_w/T_w}$, $a_2=e^{-\omega_w/T_h}$, $b_1=\omega_w\varepsilon_C/T_c$, $b_2=\omega_w\varepsilon_C/T_h$ and $p(x)$ is positive. As already mentioned, the cold heat current $\dot{\mathcal{Q}}_c(x)$ is concave and positive in the unit interval, vanishing at the boundaries $\dot{\mathcal{Q}}_c(0)=\dot{\mathcal{Q}}_c(1)=0$.

Let us take a closer look to $g(z)\equiv z^{d_c} (a_1 e^{-b_1 z} - a_2 e^{-b_2 z})$. Its analytic continuation into the complex plane has, in addition to its zeroes on the real axis, the following complex roots
\begin{equation}
z_n=1+\frac{2\pi i~T_h T_w}{\omega_w(T_w-T_h)}n\qquad n\in\mathbb{Z}\setminus\{0\}.
\label{complex_roots}\end{equation}
By applying the Hadamard factorisation theorem \cite{krantz1999handbook}, one may conveniently rewrite $g(z)$ as
\begin{equation}
g(z)= z^{d_c}~a e^{-c z}~ (1-z) \frac{\sinh\left(\frac{1-z}{2}\omega_w\frac{T_w-T_h}{T_w T_h}\right)}{(1-z)\sinh\left(\frac{\omega_w}{2}\frac{T_w-T_h}{T_w T_h}\right)},
\label{hadamard}\end{equation}
with $a,c\in\mathbb{R}^+$. In particular, $c=\omega_w\varepsilon_C(T_c+T_h)/2T_c T_h$. Back into the real axis, $\dot{\mathcal{Q}}_c(x)$ becomes
\begin{equation}
\dot{\mathcal{Q}}_c(x)= \frac{a e^{-cx}~p(x) \sinh{\epsilon(1-x)}}{(1-x)\sinh\epsilon}~ x^{d_c} (1-x),
\label{cold_heat_current4}\end{equation}
that is, we recover Eq.~\eqref{proof_sketch_1} by identifying the first factor in the r.h.s with the positive function $P(x)$. The dimensionless constant $\epsilon$ is defined as $\epsilon\equiv (T_w-T_h)\omega_w/2 T_w T_h$. From there, the performance bound $\varepsilon_* < d_c/(d_c+1)$ follows without difficulties [see Eqs.\eqref{proof_sketch_1}-\eqref{proof_sketch_3}]. Note that, the fact that $P'(x)$ has at most one zero in the unit interval follows from the concavity of $\dot{\mathcal{Q}}_c(x)$.

From Eq.~\eqref{cold_heat_current4}, one sees that taking $\varepsilon_C\rightarrow 0$ yields $P(x)\sim p(x)$, while $p(x)$ itself tends to the constant value $\dot{\mathcal{Q}}_c(x_*)$: That is, $\dot{\mathcal{Q}}_c(x)/\dot{\mathcal{Q}}_c(x_*)\sim x^{d_c}(1-x)\equiv f(x)$. To probe this limit, one can just expand $P(x)$ around $\varepsilon_C\rightarrow 0$ and take its derivative with respect to $x$ at $x=d_c/(d_c+1)$. This yields
\begin{multline}
\left. P'(x)\right\vert_{x=\frac{d_c}{d_c+1}}=\cosh{\epsilon}-\cosh\left[{
   \frac{\epsilon (1-d_c)}{d_c+1}}\right]  \\ - \frac{2 d_c \epsilon\sinh{\epsilon}}{(d_c+1)^2} +\mathcal{O}\left(\varepsilon_C^{d_c+2}\right),
\end{multline}
\newline
which is a strictly negative function of $\epsilon$. Consequently, $P'(x)\leq 0$ at $x=d_c/(d_c+1)$ for $\varepsilon_C\rightarrow 0$. Indeed, it can be seen that $P'(x)\leq 0$ $\forall~x\in(0,1)$ in this limit.
\newline

\noindent\textbf{Three-level breakup of the two-qubit fridge.} We shall finally show how the two-qubit fridge of Fig.~\ref{Fig1}(b) can be broken up into two coupled three-level masers. As a consequence, following the steps of the preceding section, one can prove that the performance bound of Eq.~\eqref{performance} also applies to the two-qubit refrigerator.

\begin{figure}[t]
	\includegraphics[width=\columnwidth]{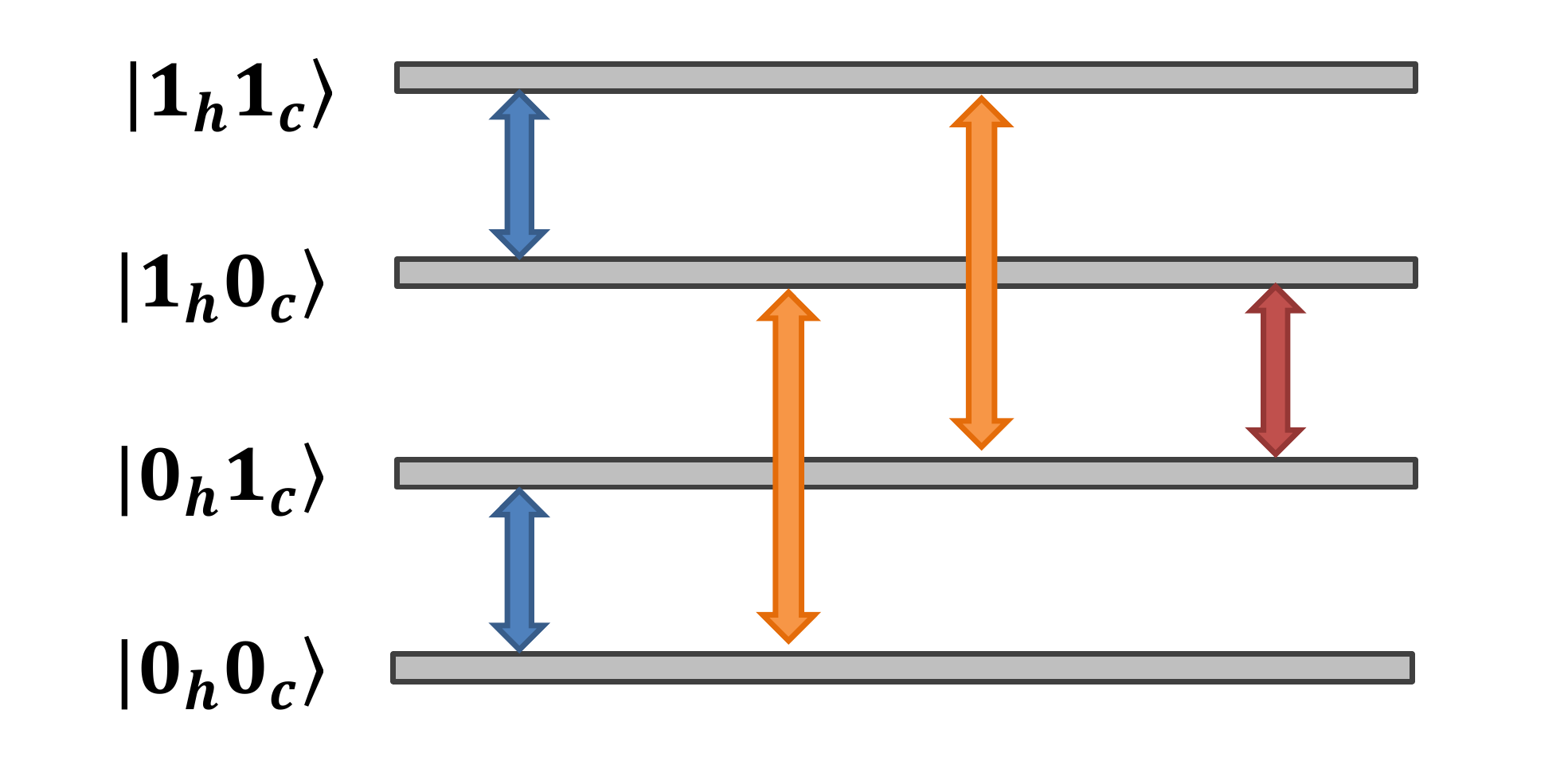}
	\caption{Level diagram of the two-qubit fridge of Fig.~\ref{Fig1}(b) with the transitions coupled to the work, hot and cold reservoirs depicted as red, orange and blue arrows respectively.}
	\label{Fig5}
\end{figure}

Let us denote the eigenstates of the working material $\hat H_\text{wm}$ by $\{\eeket{0}{0}{h}{c},\eeket{0}{1}{h}{c},\eeket{1}{0}{h}{c},\eeket{1}{1}{h}{c}\}$, with energies $\{0,\omega_c,\omega_h,\omega_h+\omega_c\}$. The cold bath couples locally to the cold qubit [see Fig.~\ref{Fig1}(b)] and thus, it only drives the transitions $\eeket{0}{0}{h}{c}\leftrightarrow\eeket{0}{1}{h}{c}$ and $\eeket{1}{0}{h}{c}\leftrightarrow\eeket{1}{1}{h}{c}$. On the other hand, the hot bath is connected to $\eeket{0}{0}{h}{c}\leftrightarrow\eeket{1}{0}{h}{c}$ and $\eeket{0}{1}{h}{c}\leftrightarrow\eeket{1}{1}{h}{c}$, while the work reservoir operates in the subspace $\eeket{0}{1}{h}{c}\leftrightarrow\eeket{1}{0}{h}{c}$ [see Fig.~\ref{Fig5}].

Intuitively, one can already see that the two-qubit fridge is comprised of two elementary three-level masers sharing the work transition. To make this statement more precise, let us write down explicitly the stationary heat currents
\begin{align*}
\dot{\mathcal{Q}}_c &= \omega_c\left(\eebra{0}{1}{h}{c}\mathcal{L}_c\hat\varrho(\infty)\eeket{0}{1}{h}{c}+\eebra{1}{1}{h}{c}\mathcal{L}_c\hat\varrho(\infty)\eeket{1}{1}{h}{c}\right) \\
\dot{\mathcal{Q}}_h &= \omega_h\left(\eebra{1}{0}{h}{c}\mathcal{L}_h\hat\varrho(\infty)\eeket{1}{0}{h}{c}+\eebra{1}{1}{h}{c}\mathcal{L}_h\hat\varrho(\infty)\eeket{1}{1}{h}{c}\right) \\
\dot{\mathcal{Q}}_w &= \omega_w\eebra{1}{0}{h}{c}\mathcal{L}_w\hat\varrho(\infty)\eeket{1}{0}{h}{c},
\end{align*}
where $\omega_w=\omega_h-\omega_c$ ($\omega_h>\omega_c$). From here, we can recover Eqs.~\eqref{components} by just defining
\begin{align*}
q_1 &= \eebra{0}{1}{h}{c}\mathcal{L}_c\hat\varrho(\infty)\eeket{0}{1}{h}{c}=-\eebra{1}{0}{h}{c}\mathcal{L}_h\hat\varrho(\infty)\eeket{1}{0}{h}{c} \\
q_2 &= \eebra{1}{1}{h}{c}\mathcal{L}_c\hat\varrho(\infty)\eeket{1}{1}{h}{c}=-\eebra{1}{1}{h}{c}\mathcal{L}_h\hat\varrho(\infty)\eeket{1}{1}{h}{c} \\
q_1 &+ q_2 = \eebra{1}{0}{h}{c}\mathcal{L}_w\hat\varrho(\infty)\eeket{1}{0}{h}{c}.
\end{align*}
As already pointed out, within the cooling window $\omega_c\leq\omega_{c,\max}$, the heat currents of each of the two three-level components $\dot{\mathcal{Q}}^{(i)}_\alpha$ behave as those of two ideal fridges, individually satisfying the first and the second laws of Eqs.~\eqref{first_law} and \eqref{second_law}. Closed formulas for $\dot{\mathcal{Q}}^{(1)}_c$ and $\dot{\mathcal{Q}}^{(2)}_c$, analogous to Eq.~\eqref{cold_heat_current2}, may be obtained from the steady-state solution of Eq.~\eqref{lindblad2}, namely
\begin{multline}
\dot{\mathcal{Q}}_c^{(i)}=\omega_c\frac{\tilde\Gamma^{i}_{\omega_c}\tilde\Gamma^{i}_{-\omega_w}\tilde\Gamma^{i}_{-\omega_c}-\tilde\Gamma^{i}_{-\omega_h}\tilde\Gamma^{i}_{\omega_w}\tilde\Gamma^{i}_{\omega_c}}{\tilde{\Delta}_{i}} \\
=\omega_c Z^i_{01}Z^i_{10}\frac{\Gamma_{\omega_h}\Gamma_{-\omega_w}\Gamma_{-\omega_c}-\Gamma_{-\omega_h}\Gamma_{\omega_w}\Gamma_{\omega_c}}{\tilde\Delta_i},
\label{component_methods_1}\end{multline}
where
\begin{subequations}
\begin{align}
\tilde\Gamma_{\omega_c}^1 &= Z^1_{01}\Gamma_{\omega_c}~,~~\tilde\Gamma_{-\omega_c}^1 = \Gamma_{-\omega_c} \\
\tilde\Gamma_{\omega_h}^1 &= Z^1_{10}\Gamma_{\omega_h}~,~~\tilde\Gamma_{-\omega_h}^1 = \Gamma_{-\omega_h} \\
\tilde\Gamma_{\omega_w}^1 &= Z^1_{10}(\Gamma_{\omega_w}+\alpha_1\Gamma_{\omega_h}\Gamma_{-\omega_c}) \\
\tilde\Gamma_{-\omega_w}^1 &= Z^1_{01}(\Gamma_{-\omega_w}+\alpha_1\Gamma_{-\omega_h}\Gamma_{\omega_c})
\end{align}
\label{component_methods_2}\end{subequations}
and
\begin{subequations}
\begin{align}
\tilde\Gamma_{\omega_c}^2 &= Z^1_{01}\Gamma_{-\omega_c}~,~~\tilde\Gamma_{-\omega_c}^2 = \Gamma_{\omega_c} \\
\tilde\Gamma_{\omega_h}^2 &= Z^2_{10}\Gamma_{-\omega_h}~,~~\tilde\Gamma_{-\omega_h}^2 = \Gamma_{\omega_h} \\
\tilde\Gamma_{\omega_w}^2 &= Z^2_{10}(\Gamma_{-\omega_w}+\alpha_2\Gamma_{-\omega_h}\Gamma_{\omega_c}) \\
\tilde\Gamma_{-\omega_w}^1 &= Z^2_{01}(\Gamma_{\omega_w}+\alpha_2\Gamma_{\omega_h}\Gamma_{-\omega_c}).
\end{align}
\label{component_methods_3}\end{subequations}
The denominator $\tilde\Delta_i$ has the same form of Eq.~\eqref{steady_state_qutrit2}, only replacing $\Gamma_{\omega_\alpha}$ with $\tilde\Gamma^i_{\omega_\alpha}$. The remaining constants in Eqs.~\eqref{component_methods_2} and \eqref{component_methods_3} are
\begin{align*}
\alpha_1 &= (\Gamma_{\omega_h}+\Gamma_{\omega_c})^{-1}~,~~\alpha_2=(\Gamma_{-\omega_h}+\Gamma_{-\omega_c})^{-1} \\
Z_{01}^ 1 &= \frac{\Gamma_{\omega_h}+\Gamma_{\omega_c}}{\Gamma_{\omega_h}+\Gamma_{\omega_c}+\Gamma_{-\omega_h}},~
Z_{10}^ 1 = \frac{\Gamma_{\omega_h}+\Gamma_{\omega_c}}{\Gamma_{\omega_h}+\Gamma_{\omega_c}+\Gamma_{-\omega_c}} \\
Z_{01}^ 2 &= \frac{\Gamma_{-\omega_h}+\Gamma_{-\omega_c}}{\Gamma_{-\omega_h}+\Gamma_{-\omega_c}+\Gamma_{\omega_h}},~
Z_{10}^ 2 = \frac{\Gamma_{-\omega_h}+\Gamma_{-\omega_c}}{\Gamma_{-\omega_h}+\Gamma_{-\omega_c}+\Gamma_{\omega_c}}.
\end{align*}
It is clear that both $\dot{\mathcal{Q}}_c^{(1)}$ and $\dot{\mathcal{Q}}_c^{(2)}$ can be cast in the generic form of Eq.~\eqref{cold_heat_current3}, so that, their associated efficiencies at maximum power must be bounded by $\varepsilon_*^{(i)}< d_c/(d_c+1)$.

\bibliographystyle{naturemag}

\section*{Acknowledgements}
The authors are grateful to K. Hovhannisyan, P. Skrzypczyk, N. Brunner, M. Huber, R. Silva, J. Goold, K. Modi, R. Kosloff, J. Rossnagel, J. G. Coello, A. Ac\'{i}n, and S. Lloyd for fruitful discussions and useful comments. This project was funded by the Spanish MICINN (Grant No. FIS2010-19998) and the European Union (FEDER), by the Canary Islands Government through the ACIISI fellowships (85\% co funded by European Social Fund), by the COST Action MP1006, by the University of Nottingham through an Early Career Research and Knowledge Transfer Award and an EPSRC Research Development Fund Grant (PP-0313/36), and by the Brazilian funding agency CAPES (Pesquisador Visitante Especial-Grant No. 108/2012).

\section*{Author contributions}
L.A.C. and J.P.P. conceived the main idea. All the authors participated in the mathematical derivations, the discussion of the results and the preparation of the manuscript.

\section*{Additional information}
The authors declare no competing financial interests. Correspondence and requests for materials should be addressed to L.A.C. (\verb"lacorrea@ull.es")

\end{document}